\documentclass[10pt,journal,compsoc]{IEEEtran}
\ifCLASSOPTIONcompsoc
  \usepackage[nocompress]{cite}
\else
  \usepackage{cite}
\fi
\ifCLASSINFOpdf
\else
\fi
\usepackage{algorithm}
\usepackage{amsmath}
\usepackage{mathrsfs}
\usepackage{amssymb}
\usepackage{mathtools}                %
\newtheorem{proposition}{Proposition}
\newtheorem{theorem}{Theorem}
\newtheorem{definition}{Definition}

\usepackage{algorithmic}

\makeatletter
\def\algbackskip{\hskip-\ALG@thistlm}
\makeatother

\ifCLASSOPTIONcompsoc
  \usepackage[caption=false,font=normalsize,labelfont=sf,textfont=sf]{subfig}
\else
  \usepackage[caption=false,font=footnotesize]{subfig}
\fi

\usepackage{acronym}
\usepackage{booktabs}
\usepackage{import}

\usepackage{pdfsync}
\usepackage{xcolor}
\colorlet{change1}{blue}

\DeclareMathOperator*{\argmin}{arg\,min} %
\DeclareMathOperator*{\argmax}{arg\,max}

\acrodef{cbmember}[CBMeMBer]{cardinality-balanced multi-object multi-Bernoulli}
\acrodef{cellmb}[cell-MB]{cell multi-Bernoulli}
\acrodef{cphd}[CPHD]{cardinalized probability hypothesis density}
\acrodef{cs}[CS]{Cauchy-Schwarz}
\acrodef{csd}[CSD]{Cauchy-Schwarz Divergence}
\acrodef{ct}[CT]{coordinated turn}
\acrodef{dddGLMB}[DD$\delta$-GLMB]{data-driven $\delta$-generalized labeled multi-Bernoulli}
\acrodef{dglmb}[$\delta$-GLMB]{$\delta$-generalized  labeled multi-Bernoulli}
\acrodef{dpgp}[DPGP]{Dirichlet process Gaussian process}
\acrodef{doc}[DOC]{distributed optimal control}
\acrodef{eer}[EER]{expected entropy reduction}
\acrodef{ekf}[EKF]{extended Kalman filter}
\acrodef{eom}[EOM]{equation of motion}
\acrodefplural{EOM}[EOMs]{equations of motion}
\acrodef{fisst}[FISST]{finite set statistics}
\acrodef{for}[FoR]{field-of-regard}
\acrodefplural{for}[FoRs]{fields-of-regard}
\acrodef{fov}[FoV]{field-of-view}
\acrodefplural{fov}[FoVs]{fields-of-view}
\acrodef{gm}[GM]{Gaussian mixture}
\acrodef{glmb}[GLMB]{generalized labeled multi-Bernoulli}
\acrodef{gospa}[GOSPA]{generalized optimal sub-pattern assignment}
\acrodef{iid}[i.i.d.]{independently and identically distributed}
\acrodef{iidc}[i.i.d.c.]{independently and identically distributed cluster}
\acrodef{imm}[IMM]{interacting multiple model}
\acrodef{jpda}[JPDA]{joint probabilistic data association}
\acrodef{jipda}[JIPDA]{joint integrated probabilistic data association}
\acrodef{jms}[JMS]{jump Markov system}
\acrodef{kl}[KL]{Kullback-Leibler}
\acrodef{kld}[KLD]{Kullback-Leibler divergence}
\acrodef{lti}[LTI]{linear time invariant}
\acrodef{lmb}[LMB]{labeled multi-Bernoulli}
\acrodef{lrfs}[LRFS]{labeled random finite set}
\acrodef{mb}[MB]{multi-Bernoulli}
\acrodef{mbm}[MBM]{multi-Bernoulli mixture}
\acrodef{mbib}[M-BIBE]{multi-Bernoulli initilization of birth targets}
\acrodef{mht}[MHT]{multiple hypothesis tracking}
\acrodef{mpc}[MPC]{model predictive control}
\acrodef{mprint}[M-PrNTT]{multi-Bernoulli probabilistic new target tracker}
\acrodef{mtt}[MTT]{multitarget tracking}
\acrodef{ONR}[ONR]{Office of Naval Research}
\acrodef{ospa}[OSPA]{optimal subpattern assignment}
\acrodef{pdf}[pdf]{probability density function}
\acrodef{peecs}[PEECS]{posterior expected error of cardinality and state}
\acrodef{pent}[PENT]{posterior expected number of targets}
\acrodef{penti}[PENTI]{posterior expected number of targets of interest}
\acrodef{pims}[PIMS]{predicted ideal measurement set}
\acrodef{pmbm}[PMBM]{Poisson multi-Bernoulli mixture}
\acrodef{pmf}[pmf]{probability mass function}
\acrodef{pomdp}[POMDP]{partially-observed Markov decision process}
\acrodef{phd}[PHD]{probability hypothesis density}
\acrodefplural{phd}[PHDs]{probability hypothesis densities}
\acrodef{radec}[RA/Dec]{right ascension / declination}
\acrodef{rfs}[RFS]{random finite set}
\acrodefplural{rfs}[RFSs]{random finite sets}
\acrodef{roi}[ROI]{region of interest}
\acrodefplural{roi}[ROIs]{regions of interest}
\acrodef{rv}[RV]{random variable}
\acrodef{smc}[SMC]{sequential Monte Carlo}
\acrodef{snr}[SNR]{signal-to-noise ratio}
\acrodef{ssa}[SSA]{space situational awareness}
\acrodef{swt}[SWT]{search-while-tracking}
\acrodef{tle}[TLE]{two-line element}
\acrodef{wami}[WAMI]{wide area motion imagery}

\newlength{\figureheight}
\newlength{\figurewidth}

\begin{document}
\title{Cell Multi-Bernoulli (Cell-MB) Sensor Control for Multi-object Search-While-Tracking (SWT)}

\author{Keith~A.~LeGrand,~\IEEEmembership{Member,~IEEE,}
        Pingping~Zhu,~\IEEEmembership{Member,~IEEE,}
        and~Silvia~Ferrari~\IEEEmembership{Senior~Member,~IEEE}%
        \thanks{Keith A.\ LeGrand and Silvia Ferrari are with the Laboratory for Intelligent Systems and Controls (LISC), Sibley School of Mechanical and Aerospace Engineering, Cornell University, Ithaca, New York, United States. Pingping Zhu is with the Department of Electrical Engineering at Marshall University, Huntington, West Virginia, United States. This work was supported in part by Office of Naval Research Grant N0014-19-1-2266, and by the Laboratory Directed Research and Development program at Sandia National Laboratories, a multi-mission laboratory managed and operated by National Technology and Engineering Solutions of Sandia, LLC, a wholly owned subsidiary of Honeywell International, Inc., for the U.S. Department of Energy’s National Nuclear Security Administration under contract DE-NA0003525.}%
}

\markboth{LeGrand \MakeLowercase{\textit{et al.}}}%
{LeGrand \MakeLowercase{\textit{et al.}}}

\maketitle

\begin{abstract}
Information-driven control can be used to develop intelligent sensors that can optimize their measurement value based on environmental feedback.
In object tracking applications, sensor actions are chosen based on the expected reduction in uncertainty also known as information gain.
Random finite set (RFS) theory provides a formalism for quantifying and estimating information gain in multi-object tracking problems.
However, estimating information gain in these applications remains computationally challenging.
This paper presents a new tractable approximation of the RFS expected information gain applicable to sensor control for multi-object search and tracking.
Unlike existing RFS approaches, the information gain approximation presented in this paper considers the contributions of non-ideal noisy measurements, missed detections, false alarms, and object appearance/disappearance.
The effectiveness of the information-driven sensor control is demonstrated through two multi-vehicle search-while-tracking experiments using real video data from remote terrestrial and satellite sensors.
\end{abstract}


\IEEEpeerreviewmaketitle

\section{Introduction}
\IEEEPARstart{M}{any} modern multi-object tracking applications involve mobile and reconfigurable sensors able to control the position and orientation of their \ac{fov} in order to expand their operational tracking capacity and improve state estimation accuracy when compared to fixed sensor systems.
By incorporating active sensor control in these dynamic tracking systems, the sensor can autonomously make decisions that produce observations with the highest information content based on prior knowledge and sensor measurements\cite{FerrariInformationDrivenPlanning21, WeiAutomaticPanTiltDgpg19, WeiInformationValueDgpg16}.
Also, the sensor \ac{fov} is able to move and cover large regions of interest, potentially for prolonged periods of time.
By expanding the autonomy and operability of sensors, however, several new challenges are introduced.
As the sensor moves and reconfigures itself, the number of objects inside the \ac{fov} changes over time.
Also, both the number of objects and the objects' states are unknown, time-varying, and subject to significant measurement errors.
As a result, existing tracking algorithms and information gain functions (e.g., \cite{FerrariInformationDrivenPlanning21, WeiAutomaticPanTiltDgpg19, WeiInformationValueDgpg16}) that assume a known number of objects and known data association, are either inapplicable or significantly degrade in performance due to measurement noise, object maneuvers, missed/spurious detections, and unknown measurement origin.

Through the use of \ac{rfs} theory, this paper formulates the multi-object information-driven control problem as a \ac{pomdp}.
Sensor actions can then be decided to maximize the expected information gain conditioned on a probabilistic information state.
Information-theoretic functionals, such as \ac{eer} \cite{CaiDemining09,ZhangComparisonInformationFunctions12}, \ac{csd} \cite{HoangCauchySchwarzPoisson15, BeardSensorControlCauchySchwarzGlmb15}, \ac{kld} \cite{GoodmanMathematicsOfDataFusion97}, and R\'{e}nyi divergence \cite{RisticSensorControlRenyi10, RisticRewardFunctionPhd11}, have been successfully used to represent sensing objectives, such as detection, classification, identification, and tracking, circumventing exhaustive enumeration of all possible outcomes.
However, \ac{rfs}-based information-theoretic sensor control policies remain computationally challenging.
Alternatively, they require simplifying assumptions that limit their applicability to vision-based \ac{swt} systems.
Tractable solutions to date employ the so-called \ac{pims} approximation \cite{MahlerProbabilisticObjectiveFunctions04}, by which sensor actions are selected based on ideal measurements with no measurement noise, false alarms, or missed detections.
This paper presents a new computationally tractable higher-order approximation called the \ac{cellmb} approximation for a restricted class of multi-object information gain functions satisfying cell-additivity constraints.
Unlike existing methods, the \ac{cellmb} approximation accounts for higher-order effects due to false alarms, missed detections, and non-Gaussian object probability distributions.

\begin{figure}[htpb]
  \centering
  \includegraphics[width=0.95\linewidth]{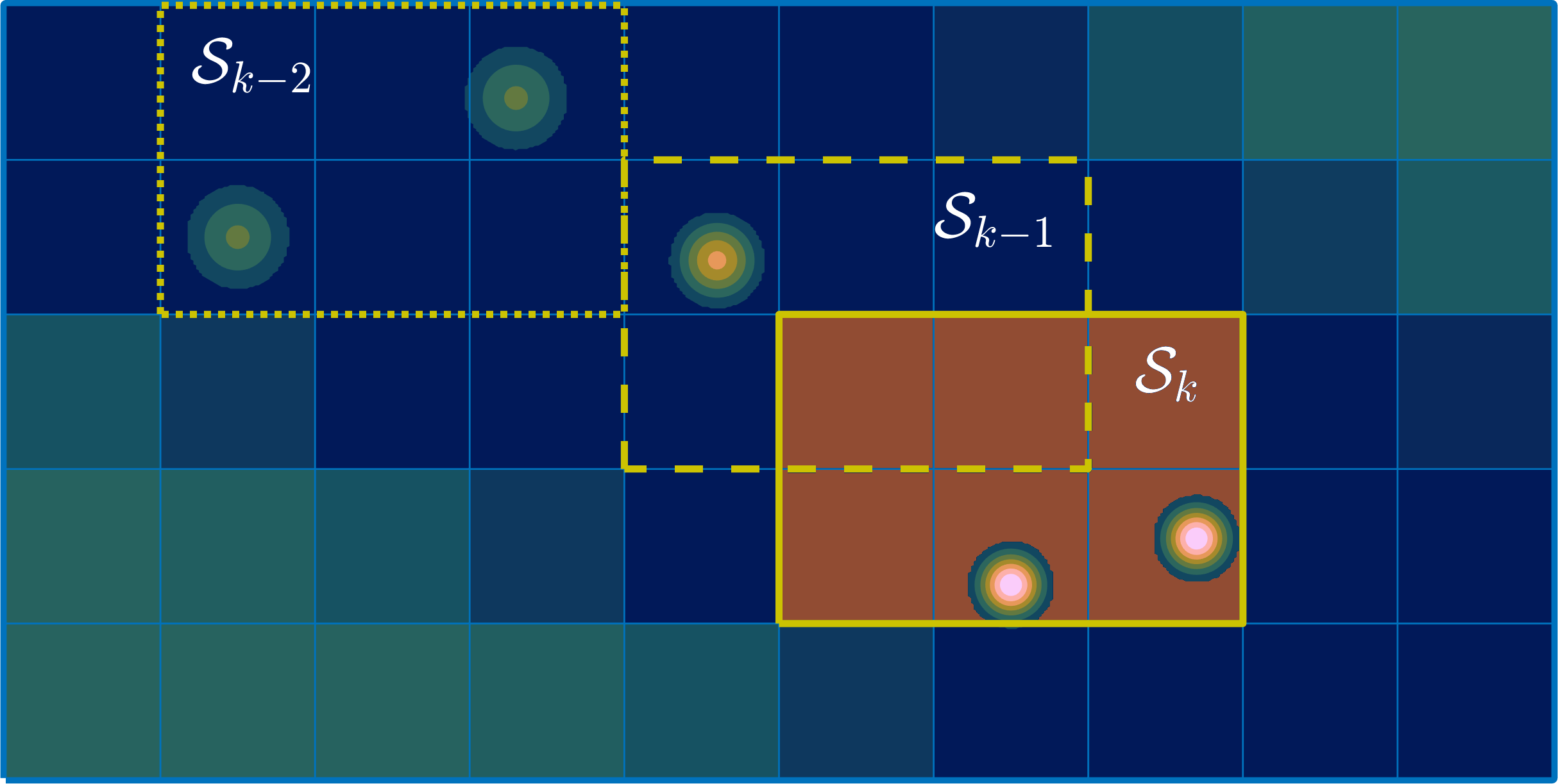}
  \includegraphics[width=0.95\linewidth]{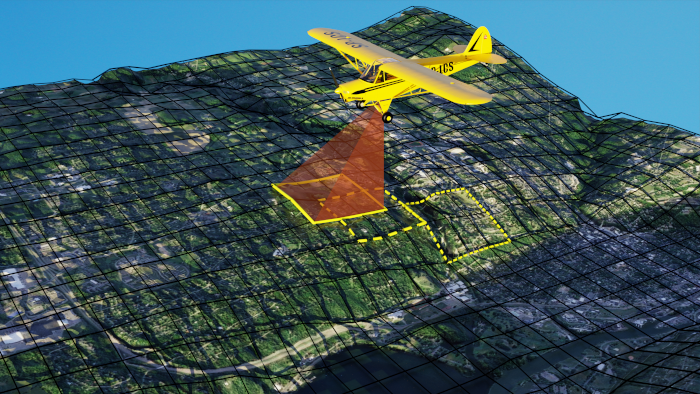}
  \caption{Conceptual image of multi-object search-while-tracking, wherein the sensor field-of-view $\mathcal{S}$ is controlled to maximize the cell multi-Bernoulli approximated information gain.}%
  \label{fig:figures/Concept}
\end{figure}
The \ac{cellmb} approximation and \ac{kld} information gain function presented in this paper also account for both discovered and undiscovered objects by enabling the efficient computation of the \ac{rfs} expectation operation.
In particular, a partially piecewise homogeneous Poisson process is used to model undiscovered objects efficiently over space and time, including in challenging settings in which objects are diffusely distributed over a large geographic region.
Prior work in \cite{OlofssonInformedPathPlanningPhd20} established a multi-agent \ac{phd}-based path planning algorithm aimed at maximizing the detection of relatively static objects.
In \cite{GehlySearchDetectTrack18}, the exploration/exploitation problem was addressed by establishing an information-theoretic uncertainty threshold for triggering pre-planned search modalities.
The occupancy grid approach in \cite{NguyenMultiobjectiveMultiagentPlanningDiscoverTrack20} was successfully implemented for tracking and discovering objects with identity-tagged observations.
A unified search and track solution was also proposed in \cite{BostromRostSensorManagementPMBM21} based on \ac{pmbm} priors and a non-information-theoretic reward.
However, these existing methods all rely on the \ac{pims} approximation and, therefore, neglect the contribution of non-ideal measurements in the prediction of information gain.
Preliminary results of this work were reported in \cite{LegrandRandomFiniteSetSensorControl21}.

The new \ac{rfs} information-driven approach presented in this paper derives a \ac{cellmb} approximation of the \ac{rfs} information gain expectation that accounts for non-ideal measurements.
A new \ac{kld} function is shown to be cell-additive and employed to represent information gain for discovered and undiscovered objects and, subsequently, is approximated efficiently using the \ac{cellmb} decomposition.
The effectiveness of this new approach is demonstrated using real video data in two distinct and challenging tracking applications involving multiple closely-spaced ground and marine vehicles maneuvering in a cluttered and remote environment.
The proposed approach is demonstrated by tracking and maintaining discovered vehicles using an optical sensor with a bounded \ac{fov}, while simultaneously searching and discovering new vehicles as they enter the surveillance region.

\section{Problem Formulation}
This paper considers an online \ac{swt} problem involving a single sensor with a bounded and mobile \ac{fov} that can be manipulated by an automatic controller, as illustrated in Fig.~\ref{fig:figures/Concept}.
The sensor objective is to discover and track multiple unidentified moving objects in a \ac{roi} that far exceeds the size of the \ac{fov}.
The objects are characterized by partially hidden states and are subject to unknown random inputs, such as driver commands, and may leave and enter the \ac{roi} at any time.
The sensor control inputs are to be optimized at every time step in order to maximize the expected reduction in track uncertainty, as well as the overall state estimation performance.

The number of objects is unknown \textit{a priori} and changes over time because objects enter and exit the surveillance region as well as, potentially, the sensor \ac{fov}.
Let $N_k$ denote the number of objects present in the surveillance region $\mathcal{W}$ at time $t_{k}$. The multi-object state $X_{k}$ is the collection of $N_{k}$ single-object states at time $t_{k}$ and is expressed as the finite set
\begin{equation}
  X_{k} = \{\mathbf{x}_{k,1}, \hdots, \mathbf{x}_{k, N_{k}}\} \, \in \, \mathcal{F}(\mathbb{X})
\end{equation}
where $\mathbf{x}_{k,i}$ is the $i$\textsuperscript{th} element of $X_{k}$ and $\mathcal{F}(\mathbb{X})$ denotes the collection of all finite subsets of the object state space $\mathbb{X}$.
Throughout this paper, single-object states are represented by lowercase letters, while multi-object states are represented by finite sets and denoted by italic uppercase letters.
Bold lowercase letters are used to denote vectors.
Spaces are represented by blackboard bold symbols, where $\mathbb{N}_{\ell}$ denotes the set of natural numbers
\begin{align}
  \mathbb{N}_{\ell} \triangleq \{1, \hdots, \ell\}
\end{align}

The multi-object measurement is the collection of $M_{k}$ single-object measurements at time $t_{k}$ and is expressed as the set
\begin{equation}
  Z_{k} = \{\mathbf{z}_{k,1}, \hdots, \mathbf{z}_{k, M_{k}} \} \, \in \, \mathcal{F}(\mathbb{Z})
\end{equation}
where $\mathbb{Z}$ denotes the measurement space.
The sensor resolution is such that single-object detections $\mathbf{z}_{k,i}$ are represented by points, e.g., a centroidal pixel, with no additional classification-quality information.
Because detections contain no identifying labels or features, the association between tracked objects and incoming measurement data is unknown.

Often in tracking, object detection may depend only a partial state $\mathbf{s}\in\mathbb{X}_{s} \subseteq \mathbb{R}^{n_{s}}$, where $\mathbb{X}_{s} \times \mathbb{X}_{v} = \mathbb{X}\subseteq \mathbb{R}^{n_{x}}$ forms the full object state space.
For example, the instantaneous ability of a sensor to detect an object may depend only on the object's relative position.
In that case, $\mathbb{X}_{s}$ is the position space, and $\mathbb{X}_{v}$ is composed of non-position states, such as object velocity.
This nomenclature is adopted throughout the paper while noting that the approach is applicable to other state definitions.

The sensor \ac{fov} is defined as a compact subset $\mathcal{S}_{k}\subset \mathbb{X}_{s}.$
Then, object detection is assumed to be random and characterized by the probability function,
\begin{equation}
  p_{D,k}(\mathbf{x}_{k}; \mathcal{S}_{k}) = 1_{\mathcal{S}_{k}} (\mathbf{s}_{k}) \cdot p_{D,k}(\mathbf{s}_{k})
\end{equation}
where the single-argument function $p_{D,k}(\mathbf{s}_{k})$ is the probability of object detection for an unbounded \ac{fov}.
When an object is detected, a noisy measurement of its state $\mathbf{x}_k$ is produced according to the likelihood function
\begin{equation}
  \mathbf{z}_k \sim g_k(\mathbf{z}_{k}|\mathbf{x}_{k})
\end{equation}
where $\mathbf{z}_k \in \mathbb{Z}$.
In addition to detections originating from true objects, the sensor produces extraneous measurements due to random phenomena, which are referred to as clutter or false alarms.
Each resolution cell (e.g., a pixel) of the sensor image plane is equally likely to produce a false alarm, and thus, the clutter process is modeled as a Poisson \ac{rfs} process with \ac{phd} $\kappa_{c}(\mathbf{z})$ \cite{BarshalomTrackingDataAssociation11}.
Further discussion on Poisson \acp{rfs} and the \ac{phd} function can be found in Section~\ref{sec:PoissonRfs}.

Let $\mathbf{u}_{k} \in \mathbb{U}_{k}$ denote the sensor control inputs that, through actuation, determine the position of the sensor \ac{fov} at time $t_{k}$, $\mathcal{S}_{k}$,
where $\mathbb{U}_{k}$ is the set of all admissible controls.
The control $\mathbf{u}_{k}$ influences both the \ac{fov} geometry, $\mathcal{S}_{k}$, and the sensor measurements, $Z_{k}$, due to varying object visibility.
Because in many modern applications the surveillance region $\mathcal{W}$ is much larger than the sensor \ac{fov}, only a fraction of the total object population can be observed at any given time.
Therefore, given the admissible control inputs $\mathbb{U}_k$, let the \ac{for} be defined as
\begin{equation}
  \label{eq:FieldOfRegardConstruction}
  \mathcal{T}_{k} \triangleq \bigcup_{\mathbf{u}_{k} \in \mathbb{U}_{k}} \mathcal{S}_{k}(\mathbf{u}_{k})
\end{equation}
and represent the composite of regions that the sensor can potentially cover (although not simultaneously) at the next time step.

Then, the sensor control problem can be formulated as an \ac{rfs} \ac{pomdp}\cite{RisticSensorControlRenyi10, KrishnamurthyPomdp16,CastanonStochasticSensorManagement07}, that includes a partially- and noisily-observed state $X_{k}$, a known initial distribution of the state $f_{0}(X_{0})$, a probabilistic transition model $f_{k|k-1}(X_{k}|X_{k-1})$, a set of admissible control actions $\mathbb{U}_{k}$, and a reward $\mathcal{R}_{k}$ associated with each control action.
At every time $k$, an \ac{rfs} multi-object tracker provides the prior $f_{k|k-1}(X_{k}|Z_{0:k-1})$ and the sensor control input is chosen so as to maximize the expected information gain, or,
\begin{align}
  \label{eq:RewardPolicyGeneral}
  \mathbf{u}_{k}^{*} = \argmax_{\mathbf{u}_{k} \in \mathbb{U}_{k}}
 \left\lbrace
   \mathrm{E}
   \left[
     \mathcal{R}_{k}(Z_{k};\, \mathcal{S}_{k}, \,
     f_{k|k-1}(X_{k}| Z_{0:k-1}), \, \mathbf{u}_{0:k-1})
   \right]
 \right\rbrace
\end{align}
where $\mathrm{E}[\cdot]$ is the expectation operator and
the functional dependence of $Z_{k}$ and $\mathcal{S}_{k}$ on $\mathbf{u}_{k}$ is omitted for brevity here but is described in \cite{MahlerAdvancesStatisticalMultitargetFusion14}.
In this paper, $\mathcal{R}_{k}$ is taken to be an information gain function, while noting that the presented results are more broadly applicable to any integrable reward function satisfying the cell-additivity constraint defined in Section~\ref{sec:InformationDrivenControl}.

A computationally tractable approximation of the expected information gain in (\ref{eq:RewardPolicyGeneral}) is derived using the new \ac{cellmb} approximation presented in Section~\ref{sec:InformationDrivenControl}.
Based on this approximation, a new sensor control policy for \ac{swt} applications is obtained in Section~\ref{sec:SWTMethodology} using a dual information gain function.
The dual information gain formulation treats discovered and undiscovered objects as separate processes,  modeling undiscovered objects as a partially piecewise homogeneous Poisson process.
By this approach, a computationally efficient sensor controller is developed for \ac{swt} over potentially large geographic regions.

\section{Background on Random Finite Sets}
\label{sec:RfsBackground}
\Ac{rfs} theory is a powerful framework for solving multi-sensor multi-object information fusion problems.
In essence, \ac{rfs} theory establishes multi-object analogs to random variables, density functions, moments, and other statistics, such that multi-object problems can be solved in a top-down fashion and with theoretic guarantees.
For readers unfamiliar with \ac{rfs} theory, \cite{VoMultitargetEncyclopedia2015} provides a gentle introduction to the subject, while \cite{MahlerStatisticalMultitargetFusion07,MahlerAdvancesStatisticalMultitargetFusion14} provide a comprehensive treatment.

An \ac{rfs} $X$ is a random variable that takes values on $\mathcal{F}(\mathbb{X})$.
A \ac{lrfs} $\mathring{X}$ is a random variable that takes values on $\mathcal{F}(\mathbb{X} \times \mathbb{L})$, where $\mathbb{L}$ is a discrete label space.
Both \ac{rfs} and \ac{lrfs} distributions can be described by set density functions, as established by Mahler's \ac{fisst} \cite{MahlerStatisticalMultitargetFusion07, MahlerAdvancesStatisticalMultitargetFusion14}.
This section reviews key \ac{rfs} concepts and notation for the Poisson \ac{rfs}, \ac{mb} \ac{rfs}, and \ac{glmb} \ac{lrfs} distributions used in this paper.
\subsection{Poisson RFS}
\label{sec:PoissonRfs}
The Poisson \ac{rfs} is fundamental to \ac{rfs} multi-object tracking due to its desirable mathematical properties and its usage in modeling false alarm and birth processes.
For example, the popular \ac{phd} filter is derived from the assumption that the multi-object state is governed by a Poisson \ac{rfs} process, which, in turn, leads to a computationally efficient tracking algorithm \cite{MahlerFirstOrderMoments03, SidenbladhParticlePhd03, VoGaussianMixtureFilter06}.

The density of a Poisson-distributed \ac{rfs} $X$ is
\begin{align}
  f(X) = e^{-N_{X}} [D]^{X}
  \label{eq:PoissonDensity}
\end{align}
where $N_X$ is the object cardinality mean, and $D(\mathbf{x})$ is the \ac{phd}, or intensity function, of $X$, which is defined on the single-object space $\mathbb{X}$.
For brevity, the multi-object exponential notation,
\begin{align}
  h^{A} \triangleq \prod_{a \in A} h(a)
\end{align}
where $h^{\emptyset}\triangleq 1$, is adopted throughout.
For multivariate functions, the dot ``$\,\cdot$\,'' denotes the argument of the multi-object exponential, e.g.:
\begin{align}
  [g(a, \cdot, c)]^{B}
  \triangleq
  \prod_{b \in B}g(a, b, c)
\end{align}

The \ac{phd} is an important statistic in \ac{rfs} theory as its integral over a set $T\subseteq\mathbb{X}$ gives the expected number of objects in that set:
\begin{equation}
  \mathrm{E} [\vert X \cap T \vert] = \int_{T} D(\mathbf{x}) \mathrm{d} \mathbf{x}
\end{equation}
The \ac{phd} of a general \ac{rfs}  $X$ is given in terms of its set density $f(X)$ as \cite{MahlerFirstOrderMoments03}
\begin{equation}
  \label{eq:PhdOfGeneralSetDensity}
  D(\mathbf{x}) = \int f(\{\mathbf{x}\} \cup X') \delta X'
\end{equation}
The integral in (\ref{eq:PhdOfGeneralSetDensity}) is a set integral, defined as
\begin{equation}
  \label{eq:FisstIntegral}
  \int f(X) \delta X \triangleq \sum_{n=0}^{\infty} \frac{1}{n!}
  \int f(\{\mathbf{x}_{1}, \hdots, \mathbf{x}_{n}\})
  \mathrm{d} \mathbf{x}_{1} \cdots \mathrm{d} \mathbf{x}_{n}
\end{equation}
The set integral is a fundamental construct of \ac{rfs} theory and enables the direct translation of the Bayes' filter recursion to the multi-object setting, as shown in~\cite{MahlerStatisticalMultitargetFusion07} and discussed in Section~\ref{sec:MultiObjectFiltering}.
Set integration via~(\ref{eq:FisstIntegral}) also presents practical challenges, as exact computation is rarely possible due to the infinite summation of nested multivariate integrals required.
This challenge is a key motivation of the tractable \ac{cellmb} approximation introduced in Section~\ref{sec:InformationDrivenControl}.

\subsection{Multi-Bernoulli RFS}
In an \ac{mb} distribution, a given object's existence is modeled as a Bernoulli random variable and specified by a probability of existence.
As such, the \ac{mb} \ac{rfs} can accurately model a variety of multi-object processes when the true existence of objects is unknown and subject to change.
The density of an \ac{mb} distribution is \cite[p.\ 102]{MahlerAdvancesStatisticalMultitargetFusion14}
\begin{align}
  f(X)
  =
  \left[
    1 - r^{(\cdot)}
  \right]^{\mathbb{N}_{M}}
  \sum\limits_{\mathclap{\vphantom{\big[}1 \leq i_{1} \neq \cdots \neq i_{n} \leq M}}
    \quad \quad
  \left[
    \dfrac{
      r^{i_{(\cdot)}} p^{i_{(\cdot)}}(\mathbf{x}_{(\cdot)})
    }{
      1 - r^{i_{(\cdot)}}
    }
  \right]^{\mathbb{N}_{n}}
  \label{eq:MbRfsDensity}
\end{align}
where $n=|X|$,  $M$ is the number of \ac{mb} components and maximum possible object cardinality, $r^{i}$ is the probability that the $i$\textsuperscript{th} object exists, and $p^{i}(\mathbf{x})$ is the single-object state probability density of the $i$\textsuperscript{th} object if it exists.
Given a MB distribution with density (\ref{eq:MbRfsDensity}), its \ac{phd} is given by
\begin{equation}
  D(\mathbf{x}) = \sum_{j=1}^{M} r^{j} p^{j}(\mathbf{x})
  \label{eq:PhdOfMbRfsDensity}
\end{equation}

\subsection{GLMB RFS}
The density of a \ac{glmb} distribution proposed in \cite{VoLabeledRfsGlmbFilter14} is given by
\begin{align}
  \label{eq:GlmbDensity}
  \mathring{f} (\mathring{X})
  =
  \Delta (\mathring{X})
  \sum\limits_{\xi \in \Xi}
  w^{(\xi)} (\mathcal{L} (\mathring{X}))
  [p^{(\xi)}]^{\mathring{X}} \,,
\end{align}
where $\Xi$ is a discrete space, and where each $\xi\in\Xi$ represents a history of measurement association maps, each $p^{(\xi)}(\cdot,\ell)$ is a probability density on $\mathbb{X}$, and each weight $w^{(\xi)}$ is non-negative with
\begin{equation*}
  \sum\limits_{(I,\xi)\in \mathcal{F}(\mathbb{L})\times\Xi} w^{(\xi)}(I)=1
\end{equation*}
The label of a labeled state $\mathring{x}$ is recovered by $\mathcal{L}(\mathring{x})$, where $\mathcal{L} : \mathbb{X} \times \mathbb{L} \mapsto \mathbb{L}$ is the projection defined by $\mathcal{L}((\boldmath{x}, \ell)) \triangleq \ell$.  Similarly, for \acp{lrfs}, $\mathcal{L}(\mathring{X}) \triangleq \{\mathcal{L}(\mathring{x}) : \mathring{x} \in \mathring{X}\}$.  The distinct label indicator  $\Delta(\mathring{X})=\delta_{(|\mathring{X}|)}(|\mathcal{L}(\mathring{X})|)$ ensures that only sets with distinct labels are considered.

\subsection{Multi-Object Filtering}
\label{sec:MultiObjectFiltering}
Online estimation of the multi-object state is performed using the data-driven \ac{glmb} filter, which provides the Bayes-optimal solution of the measurement-driven Bayes filter recursion \cite{LegrandDataDrivenGlmb18}:
\begin{align}
  \label{eq:LabeledBayesFilterPredict}
  \mathring{f}_{\mathrm{p}}(\mathring{X}_{\mathrm{p},k}|Z_{0:k-1})
  &=
  \int \mathring{f}(\mathring{X}_{\mathrm{p},k}|\mathring{X}_{k-1}) \mathring{f}(\mathring{X}_{k-1}| Z_{0:k-1}) \delta \mathring{X}_{k-1} \\
  \label{eq:DataDrivenLabeledBayesFilterUpdate}
  \mathring{f}(\mathring{X}_{k}|Z_{0:k})
  &=
  \frac{g(Z_{k} | \mathring{X}_{k}) \mathring{f}_{\mathrm{p}}(\mathring{X}_{\mathrm{p},k}|Z_{0:k-1}) \mathring{f}_{\mathrm{b}}(\mathring{X}_{\mathrm{b},k})}{\int g(Z_{k} | \mathring{X}) \mathring{f}_{\mathrm{p}}(\mathring{X}_{\mathrm{p},k}|Z_{0:k-1})
  \mathring{f}_{\mathrm{b}}(\mathring{X}_{\mathrm{b},k})\delta \mathring{X}}
\end{align}
The function time indices have been suppressed for brevity, and $\mathring{f}_{\mathrm{p},k}(\mathring{X}_{\mathrm{p},k})$ and $\mathring{f}_{\mathrm{b},k}(\mathring{X}_{\mathrm{b},k})$ denote the density of persisting and birth objects, respectively, where the joint state $\mathring{X}_{k} = \mathring{X}_{\mathrm{p},k} \cup \mathring{X}_{\mathrm{b},k}$.
The accent ``$\mathring{\,\,\,}$'' is used to distinguish labeled states and functions from their unlabeled equivalents, where a state's label is simply a unique number or tuple to distinguish it from the states of other objects and associate track estimates over time.
$\mathring{f}_{k|k-1}(\mathring{X}_{\mathrm{p},k}|\mathring{X}_{k-1})$ is the multi-object transition density, $g_{k}(Z_{k}| \mathring{X}_{k})$ is the multi-object measurement likelihood function, and $g_{k}$ is used to denote both the single-object and multi-object measurement likelihood function.
The nature of the likelihood function can be easily determined from its arguments.

\subsection{Kullback-Leibler Divergence}
\begin{figure}[htpb]
  \centering
  \includegraphics[width=\linewidth]{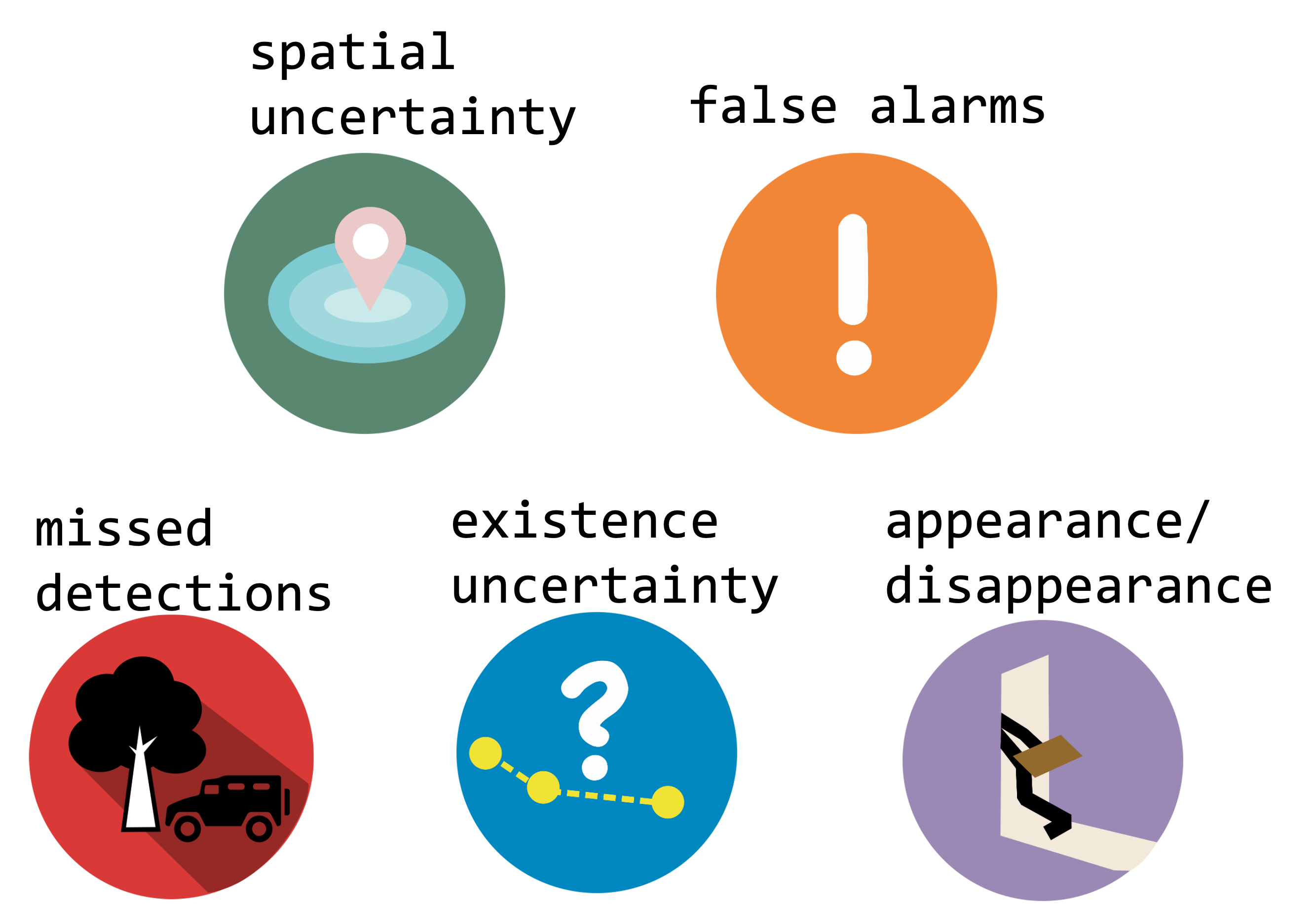}
  \caption{Graphical representation of important \ac{swt} considerations that are elegantly encapsulated in a compact and abstract \ac{rfs} information gain function.}  \label{fig:RfsInformationDrivenControlSchematic}
\end{figure}

Like single-object distributions, the similarity of \ac{rfs} distributions may be measured by the \ac{kld}.
Let $f_{1}$ and $f_{0}$ be integrable set densities where $f_{1}$ is absolutely continuous with respect to $f_{0}$.
Then, the \ac{kld} is \cite[p.~206]{GoodmanMathematicsOfDataFusion97}
\begin{equation}
  \label{eq:KullbackLeiblerDivergenceRfs}
  I(f_{1};f_{0}) = \int f_{1}(Y) \log \left(\frac{f_{1}(Y)}{f_{0}(Y)}\right) \delta Y
\end{equation}
Further simplification is possible if $f_{0}$ and $f_{1}$ are Poisson with respective \acp{phd} $D_{0}$ and $D_{1}$, in which case
\begin{equation}
  \label{eq:KldOfTwoPoisson}
  I(f_{1}; f_{0}) = N_{0} - N_{1} + \int D_{1}(\mathbf{y}) \cdot \log \left(\frac{D_{1}(\mathbf{y})}{D_{0}(\mathbf{y})}\right) \mathrm{d} \mathbf{y}
\end{equation}
where $N_{0}=\int D_{0}(\mathbf{y}) \mathrm{d} \mathbf{y}$ and $N_{1}=\int D_{1}(\mathbf{y}) \mathrm{d} \mathbf{y}$.
Importantly, when $f_{0}$ and $f_{1}$ represent prior and posterior densities, respectively, the \ac{kld} is a measure of information gain.
By formulating control policies based on \ac{rfs} divergence measures, the complicated effects of spatial uncertainty, false alarms, missed detections, existence uncertainty, and object appearance/disappearance are elegantly captured in a compact and abstract objective, as depicted in Fig.~\ref{fig:RfsInformationDrivenControlSchematic} and described in the following section.

\section{Information-Driven Control}
\label{sec:InformationDrivenControl}
The objective of information-driven control is to maximize the value of the information  gained by future measurements before they are known to the sensor.
The expected information gain, therefore, can be obtained by marginalizing over the set $Z_{k}$, using an available measurement model.
Then, the expected information gain obtained at the \emph{next} time step can be obtained from the set integral
\begin{equation}
  \label{eq:RewardExpectation}
  \mathrm{E}[\mathcal{R}_{k}] = \int \mathcal{R}_{k}(Z_{k}; \cdot) f(Z_{k}) \delta Z_{k}
\end{equation}
where $f(Z_{k})$ is the predicted measurement density conditioned on past measurements.
In general, direct evaluation of (\ref{eq:RewardExpectation}) is computationally intractable due to the infinite summation of nested single-object integrals (see (\ref{eq:FisstIntegral})).
Furthermore, each integrand evaluation encompasses a multi-object filter update and subsequent divergence computation.
As such, principled approximations are needed for tractable computation of the expected information gain.

\subsection{The Cell-MB Distribution}
A new approximation of \ac{rfs} density functions is presented in this section and, then, used to obtain the information gain expectation.
We refer to this as the \ac{cellmb} approach, which approximates an arbitrary measurement density as an \ac{mb} density with existence probabilities and single-object densities derived from a cell-decomposition of the measurement space.
\begin{definition}
  Consider the tessellation of the space $\mathbb{Y}$ into $P$ disjoint subspaces, or cells, as
\begin{equation}
  \mathbb{Y} = \overset{1}{\mathbb{Y}} \uplus \cdots \uplus \overset{P}{\mathbb{Y}}
  \label{eq:YCellDecomp}
\end{equation}
Given the cell-decomposition (\ref{eq:YCellDecomp}), the \ac{rfs}
\begin{equation*}
  Y = \{\mathbf{y}_{1}, \hdots, \mathbf{y}_n\}
\end{equation*}
is considered to be cell-MB if it is distributed according to the density
\begin{align}
  \label{eq:CellMbDensity}
  f(Y) &= \Delta(Y, \mathbb{Y}) \left[1 - r^{(\cdot)}\right]^{\mathbb{N}_{P}}\nonumber\\
       &\quad \cdot
   \sum_{1\leq j_{1} \neq \cdots \neq j_{n} \leq P}
   \left[\frac{r^{j_{(\cdot)}} p^{j_{(\cdot)}}(\mathbf{y}_{(\cdot)})}{1 - r^{j_{(\cdot)}}}\right]^{\mathbb{N}_{n}}
\end{align}
where
\begin{equation}
  \label{eq:CellSupportRestriction}
  \Delta(Y, \mathbb{Y}) \triangleq
  \begin{cases}
    1 & |Y\cap \overset{j}{\mathbb{Y}}| \leq 1 \, \forall \, j \in \{1, \hdots, P\} \\
  0 & \textrm{otherwise}
  \end{cases}
\end{equation}
and
\begin{equation}
  \int_{\overset{j}{\mathbb{Y}}} p^j (\mathbf{y}) \mathrm{d} \mathbf{y} = 1 , \qquad j=1, \ldots, P
\end{equation}
\end{definition}

Note that the \ac{cellmb} distribution is a special case of the \ac{mb} distribution in which the probability of more than one object occupying the same cell is zero.

In \cite{NussRfsDynamicOccupancyGrid18}, a collection of Bernoulli distributions was defined over an occupancy grid by integration of the \ac{phd} for dynamic map estimation applications.
Inspired by \cite{NussRfsDynamicOccupancyGrid18}, in this paper, a general \ac{cellmb} approximation is developed for an arbitrary density and appropriate cell-decomposition.
We show that the best \ac{cellmb} approximation, as defined by \ac{kld} minimization, has a matching \ac{phd} and cell weights equal to the expected number of objects in each cell, as summarized by the following proposition.

\begin{figure}[htpb]
  \centering
  \includegraphics[width=0.98\linewidth]{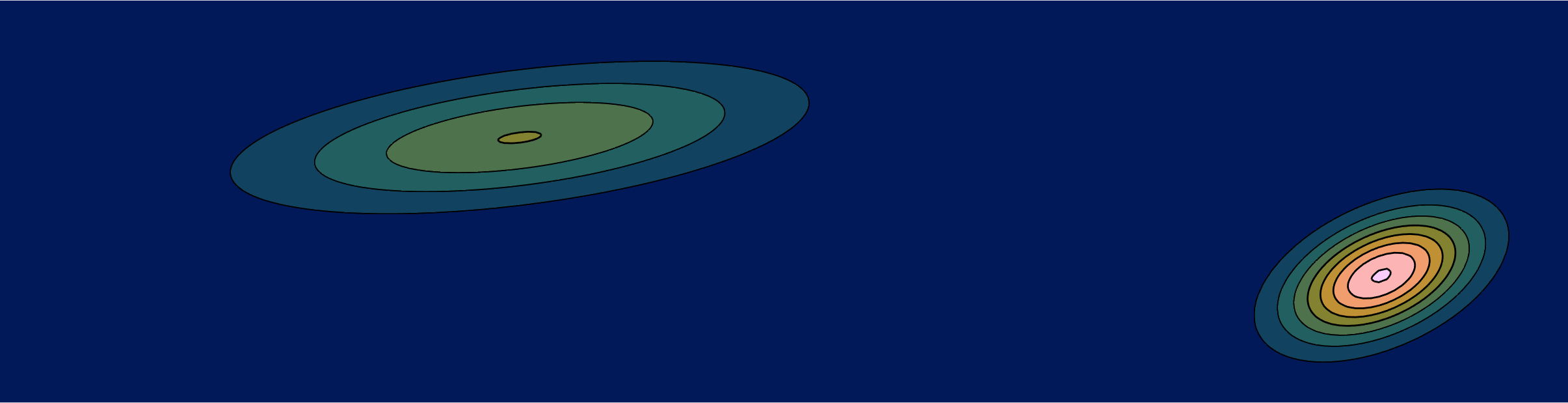}\\[1em]
  \ifx\undefined\usetikz
  \includegraphics{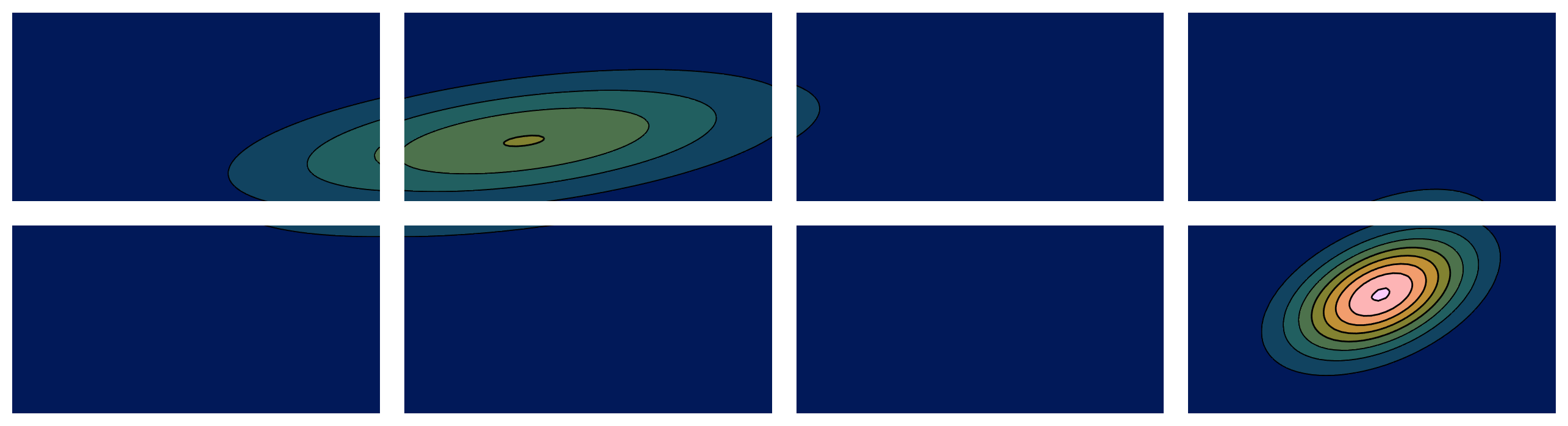}
  \else
    \setlength{\figurewidth}{0.98\linewidth}
    \import{figures/}{FigExampleMeasurementDensityMosaic.tikz}
  \fi
  \caption{The PHD of an RFS distribution $f(Y)$ (top) and its cell-MB approximation (bottom), where each cell $j$ has a corresponding probability of existence $r^{j}$ and spatial density $p^{j}(\mathbf{y})$.}%
  \label{fig:FigExampleMeasurementDensityMosaic}
\end{figure}

\begin{proposition}
  Let $f(Y)$ be an arbitrary set density with \ac{phd} $D(\mathbf{y})$ and $\overset{1}{\mathbb{Y}} \uplus \cdots \uplus \overset{P}{\mathbb{Y}}$ be a cell-decomposition of space $\mathbb{Y}$ such that
\begin{equation}
  \int_{\overset{j}{\mathbb{Y}}} D(\mathbf{y}) \mathrm{d} \mathbf{y} \leq 1,  \quad j=1,\ldots,P
\end{equation}
  If $\bar{f}(Y)$ is a \ac{cellmb} distribution over the same cell-decomposition with parameters $\{r^{j}, p^{j}\}_{j=1}^{P}$, the \ac{kld} between $f(Y)$ and $\bar{f}(Y)$ is minimized by parameters
\begin{align}
  r^{j} &= \int 1_{\overset{j}{\mathbb{Y}}}(\mathbf{y}) D(\mathbf{y}) \mathrm{d} \mathbf{y}\\
  p^{j}(\mathbf{y}) &= \frac{1}{r^{j}} 1_{\overset{j}{\mathbb{Y}}}(\mathbf{y}) D(\mathbf{y})
\end{align}
\end{proposition}
The proof is provided in Appendix~\ref{app:KldMinimizingCellMb}.
An example $2\times4$ cell-decomposition and corresponding \ac{cellmb} approximation is shown in Fig.~\ref{fig:FigExampleMeasurementDensityMosaic} for illustration.
As shown, the \ac{cellmb} approximation has a matching \ac{phd} surface, but within each cell $j$, the spatial density $p^{j}(\mathbf{y})$ is confined to the cell support $\overset{j}{\mathbb{Y}}$.
When applied to the predicted measurement density, the \ac{cellmb} approximation results in a simplified multi-object expectation for a restricted class of information gain functions, as described in the following subsection.

\subsection{Information Gain Expectation: Cell-MB}
In order to reduce the computational complexity associated with the set integral in (\ref{eq:RewardExpectation}), this subsection shows that the multi-object information gain expectation simplifies to a finite sum involving only single-object integrals, assuming the measurement is \ac{cellmb} distributed and the information gain function in (\ref{eq:RewardExpectation}) is cell-additive, as defined in this subsection.

Given the \ac{fov} $\mathcal{S}\subset \mathbb{X}_{s}$, let
\begin{align}
  \label{eq:FovCellDefinition}
  \overset{j}{\mathcal{S}} \triangleq \mathcal{S} \cap \overset{j}{\mathbb{X}}_{s}
\end{align}
Furthermore, assume that position state cells do not overlap at the \ac{fov} bounds, such that each position state cell $\overset{j}{\mathbb{X}}_{s}$ is either wholly included in or wholly excluded by $\mathcal{S}$:
\begin{equation}
  \label{eq:GridsAreAligned}
  \overset{j}{\mathbb{X}}_{s} \setminus \overset{j}{\mathcal{S}} = \emptyset \quad \forall \, \overset{j}{\mathcal{S}} \neq \emptyset
\end{equation}

This assumption is without loss of generality, as any violation is resolved by subdividing a cell $\overset{j}{\mathbb{X}}_{s}$ into two new cells $\overset{j}{\mathbb{X}}_{s}\cap \overset{j}{\mathcal{S}}$ and $\overset{j}{\mathbb{X}}_{s} \setminus \overset{j}{\mathcal{S}}$.

Then, the cell-additivity condition can be defined as follows.

\begin{definition}
  \label{def:CellAdditivity}
 Given a decomposition $\overset{1}{\mathbb{Z}} \uplus \cdots \uplus \overset{P}{\mathbb{Z}}$ of space $\mathbb{Z}$, the information gain function $\mathcal{R}_{k}(\cdot)$ is cell-additive if
  \begin{equation}
    \mathcal{R}_{k}(Z_{k}; \mathcal{S}_{k}) = \sum_{j=1}^{P} \mathcal{R}_{k}(Z_{k} \cap \overset{j}{\mathbb{Z}}; \overset{j}{\mathcal{S}_{k}})
    \label{eq:RewardAdditiveOverZ}
  \end{equation}
\end{definition}

\begin{theorem}
  \label{thm:CellMbExpectation}
  Let $Z_{k}$ be distributed according to the \ac{cellmb} density $f(Z_{k})$ with parameters $\{r^{j},p^{j}\}_{j=1}^{P}$ and the cell-decomposition
  \begin{equation}
    \mathbb{Z} = \overset{1}{\mathbb{Z}} \uplus \cdots \uplus \overset{P}{\mathbb{Z}}
    \label{eq:ZCellDecomp}
  \end{equation}
  If the information gain function $\mathcal{R}_{k}(\cdot)$ is integrable and cell-additive (Def.~\ref{def:CellAdditivity}), then the expected information gain is
  \begin{align}
    \mathrm{E}[\mathcal{R}_{k}]
    &=
    \sum_{j=1}^{P}
      \mathcal{R}_{k}(\emptyset; \overset{j}{\mathcal{S}}_{k})
      \left(1 - r^{j}\right)
      +
      \hat{\mathcal{R}}_{\mathrm{z},k}^{j}
      \cdot
      r^{j}
    \label{eq:ExpectedRewardSum}
  \end{align}
  where
  \begin{equation}
    \label{eq:ExpectedSingleSensorRewardConditionedOnSingleMeas}
    \hat{\mathcal{R}}_{\mathrm{z},k}^{j} \triangleq
    \int_{\overset{j}{\mathbb{Z}}} \mathcal{R}_{k}(\{\mathbf{z}\}; \overset{j}{\mathcal{S}_{k}}) p^{j}(\mathbf{z}) \mathrm{d} \mathbf{z}
  \end{equation}
\end{theorem}
Proof of Theorem~\ref{thm:CellMbExpectation} is given in Appendix~\ref{app:CellMbExpectationProof}.

\textit{Remark:} In (\ref{eq:RewardAdditiveOverZ}), (\ref{eq:ExpectedRewardSum}), and (\ref{eq:ExpectedSingleSensorRewardConditionedOnSingleMeas}), the auxiliary information gain arguments are suppressed for brevity and to highlight the structure of the \ac{cellmb} approximation.

The remainder of this paper considers information gain functions satisfying the cell-additivity constraint of (\ref{eq:RewardAdditiveOverZ}), such as the \ac{phd} filter update based \ac{kld} information gain.
Note that adopting the \ac{phd} filter for estimating the information gain does not require using it for multi-object tracking.
Given an arbitrary \ac{rfs} prior density $f_{k|k-1}(X)$ and its \ac{phd} $D_{k|k-1}(\mathbf{x})$, the \ac{phd} filter update based (hereon abbreviated as ``\ac{phd}-based'') \ac{kld} information gain is
\begin{align}
  \label{eq:KldPhdPseudolikelihood}
  \mathcal{R}_{k}&(Z; \mathcal{S}, D_{k|k-1})
  = \int_{\mathbb{X}} D_{k|k-1}(\mathbf{x}) \\
  &\qquad \times
  \{
    1 - L_{Z}(\mathbf{x}; \mathcal{S}) + L_{Z}(\mathbf{x}; \mathcal{S})
    \log[L_{Z}(\mathbf{x};\mathcal{S})]
  \}
  \mathrm{d} \mathbf{x}\nonumber
\end{align}
where the pseudo-likelihood function
\begin{align}
  \label{eq:PseudolikelihoodFunction}
  L_{Z}(\mathbf{x};\mathcal{S}) &= 1 - p_{D}(\mathbf{x}; \mathcal{S}) \\
  &\quad +
  \sum_{\mathbf{z}\in Z}
  \frac{p_{D}(\mathbf{x}; \mathcal{S}) \cdot g(\mathbf{z}|\mathbf{x})}{\kappa_{c}(\mathbf{z}) + \int p_{D}(\mathbf{x}; \mathcal{S}) g(\mathbf{z}|\mathbf{x}) D_{k|k-1}(\mathbf{x}) \mathrm{d} \mathbf{x}} \nonumber
\end{align}
is adopted from \cite[p.~193]{MahlerAdvancesStatisticalMultitargetFusion14}.
The following proposition establishes that (\ref{eq:KldPhdPseudolikelihood}) is cell-additive for appropriate cell-decompositions.
\begin{proposition}
  \label{prp:PhdKldIsAdditive}
Assume there exists a joint decomposition
\begin{equation}
  \mathbb{Z} = \overset{1}{\mathbb{Z}} \uplus \cdots \uplus \overset{P}{\mathbb{Z}} \,,
  \qquad
  \mathbb{X} = \overset{1}{\mathbb{X}} \uplus \cdots \uplus \overset{P}{\mathbb{X}}
\end{equation}
such that (\ref{eq:GridsAreAligned}) is satisfied and assume that an
object in cell~$\overset{j}{\mathbb{X}}$
can only generate measurements within its corresponding measurement cell~$\overset{j}{\mathbb{Z}}$; i.e.:
\begin{align}
  \label{eq:NoCellOverlap}
  D_{k|k-1}(\mathbf{x}) g_{k}(\mathbf{z}| \mathbf{x}) = 0
  \qquad \forall \,
  \, \mathbf{x} \in \overset{j}{\mathbb{X}},
  \, \mathbf{z} \in \overset{j'}{\mathbb{Z}}, j \neq j'
\end{align}
Then, the \ac{phd}-based \ac{kld} is cell-additive:
\begin{equation}
  \mathcal{R}_{k}(Z; \mathcal{S}, D_{k|k-1})
  =
  \sum_{j=1}^{P}
  \mathcal{R}_{k}(Z \cap \overset{j}{\mathbb{Z}}; \overset{j}{\mathcal{S}}, D_{k|k-1})
\end{equation}
\end{proposition}
Proof of Proposition~\ref{prp:PhdKldIsAdditive} is provided in Appendix~\ref{app:PhdKldIsAdditiveProof}.
Proposition~\ref{prp:PhdKldIsAdditive} establishes that for appropriate cell-decompositions, the \ac{phd}-based \ac{kld} for a given \ac{fov} is equivalent to the sum of \ac{phd}-based \ac{kld} information gains for smaller ``virtual'' \acp{fov}.
Perfect cell-additivity requires satisfying (\ref{eq:NoCellOverlap}), which, in turn, implies that an object in cell $\overset{i}{\mathbb{X}}$ does not generate a measurement in $\overset{j}{\mathbb{Z}}$ for $i\neq j$.
In general, violations of (\ref{eq:NoCellOverlap}) are tolerable and result in approximation errors that are negligible in comparison to the stochastic variations in the actual information gain.
Furthermore, these simplifying assumptions need not be satisfied by the multi-object tracker.

The \ac{cellmb} approximation accounts for the potential information gain of non-ideal measurements, which may include missed detections, clutter, and measurements originating from new objects.
The latter case is particularly important for the search for undiscovered objects, as is shown in the following section.

\section{Search-While-Tracking Sensor Control}
\label{sec:SWTMethodology}
This section presents a dual information gain function and associated sensor control policy that takes into account both discovered and undiscovered objects.
The information gain function proposed in Section~\ref{sec:dual_reward} balances the competing objectives of object search and tracking by means of a unified information-theoretic framework.
Sections~\ref{sec:discovered_object_reward_expectation} and~\ref{sec:undiscovered_object_reward_expectation} derive the expected information gain functions for discovered and undiscovered objects, respectively, the combination of which is maximized by the sensor control policy in Section~\ref{sec:FieldOfRegard}.
Sections~\ref{sec:undiscovered_object_prediction_and_update} and~\ref{sec:discovered_object_tracking} describe multi-object filters for recursive estimation of the undiscovered and discovered object densities, respectively, and the overall \ac{swt} algorithm is summarized in Section~\ref{sec:algorithm_overview}.

\subsection{Dual Information Gain Function}%
\label{sec:dual_reward}
Separate density parameterizations for discovered and undiscovered objects are employed such that their unique characteristics may be leveraged for computational efficiency.
Let $X_{u, k}\in \mathcal{F}(\mathbb{X})$ be the state of objects that were not detected during steps $0,\ldots,k-1$ and $X_{d,k} \in \mathcal{F}(\mathbb{X})$ be the state of objects detected prior to $k$.
Denote by $Z_{u, k}$, $Z_{d, k}$, and $Z_{c, k}$ the detections generated by $X_{u,k}$, $X_{d,k}$, and clutter, respectively.
Let $V_{k}\triangleq Z_{d,k} \cup Z_{c,k}$ and $W_{k}\triangleq Z_{u,k} \cup Z_{c,k}$.
Then, the sensor control policy is defined in terms of the dual information gain as
\begin{align}
  \mathbf{u}_{k}^{*} =
  \argmax_{\mathbf{u}_{k} \in \mathbb{U}_{k}}
  \bigg\lbrace
    &\mathrm{E}[\mathcal{R}_{k}^{d}(V_{k};\mathcal{S}_{k}(\mathbf{u}_{k}))]\nonumber\\
    &+
    \mathrm{E}[\mathcal{R}_{k}^{u}(W_{k};\mathcal{S}_{k}(\mathbf{u}_{k}))]
  \bigg\rbrace
\end{align}
where
\begin{align}
  \mathcal{R}_{k}^{d}(\cdot;\cdot)&=\mathcal{R}_{k}(\cdot;\cdot,D_{d,k|k-1}) \\
  \mathcal{R}_{k}^{u}(\cdot;\cdot)&=\mathcal{R}_{k}(\cdot;\cdot,D_{u,k|k-1})
\end{align}
are used for brevity, and $D_{d,k|k-1}$ and $D_{u,k|k}$ are the prior \acp{phd} of discovered and undiscovered objects, respectively.
The individual information gain expectations for discovered and undiscovered objects are derived in the following subsections.

\subsection{Expected Information Gain of Discovered Objects}%
\label{sec:discovered_object_reward_expectation}
If $f_{k|k-1}(V_{k})$ is \ac{cellmb} with parameters $\{r_{\mathrm{v}}^{j}, p_{\mathrm{v}}^{j}\}_{j=1}^{P}$, then from Theorem~\ref{thm:CellMbExpectation} it follows that
\begin{align}
  \mathrm{E}[\mathcal{R}_{k}^{d}]
  &=
  \sum_{j=1}^{P}
    \mathcal{R}_{k}^{d}(\emptyset; \overset{j}{\mathcal{S}}_{k})
    \left(1 - r_{\mathrm{v}}^{j}\right)%
  +
    \hat{\mathcal{R}}_{\mathrm{v},k}^{d,j}(\overset{j}{\mathcal{S}}_{k})
    \cdot
    r_{\mathrm{v}}^{j}
\end{align}
where
\begin{align}
  \label{eq:DiscoveredObjectConditionalReward}
  \hat{\mathcal{R}}_{\mathrm{v},k}^{d,j}(\overset{j}{\mathcal{S}}) &\triangleq
  \int_{\overset{j}{\mathbb{Z}}} \mathcal{R}_{k}^{d}(\{\mathbf{z}\}; \overset{j}{\mathcal{S}}) p_{\mathrm{v}}^{j}(\mathbf{z}) \mathrm{d} \mathbf{z} \\
  \label{eq:DiscoveredObjectMeasurementCellMb_r}
  r_{\mathrm{v}}^{j}(\mathcal{S}) &= \int 1_{\overset{j}{\mathbb{Z}}}(\mathbf{z}) D_{\mathrm{v},k|k-1}(\mathbf{z}; \mathcal{S}) \mathrm{d} \mathbf{z}\\
  \label{eq:DiscoveredObjectMeasurementCellMb_p}
  p_{\mathrm{v}}^{j}(\mathbf{z}; \mathcal{S}) &= \frac{1}{r_{\mathrm{v}}^{j}} 1_{\overset{j}{\mathbb{Z}}}(\mathbf{z}) D_{\mathrm{v},k|k-1}(\mathbf{z}; \mathcal{S})
\end{align}
The multi-object tracker provides the prior \ac{glmb} density $\mathring{f}_{\mathrm{p},k|k-1}(\mathring{X}_{\mathrm{p},k}|Z_{0:k})$, from which the discovered object \ac{phd} is obtained as
\begin{equation}
  \label{eq:PhdDiscoveredObjectsFromGlmb}
  D_{d,k|k-1}(\mathbf{x}) =
  \sum_{(I,\xi) \in \mathcal{F}(\mathbb{L}) \times \Xi} \sum_{\ell \in I}
  w^{(\xi)}(I) p^{(\xi)}(\mathbf{x}, \ell)
\end{equation}
The \ac{phd} $D_{\mathrm{v},k|k-1}$ can be obtained from the predicted measurement density $f_{k|k}(V_{k})$ through application of (\ref{eq:PhdOfGeneralSetDensity}).
From the prior \ac{glmb} density,
\begin{equation}
  f_{k|k-1}(V_{k}) = \int g_{k}(V_{k}|\mathring{X}) \mathring{f}_{k|k-1}(\mathring{X}) \delta \mathring{X}
\end{equation}
Given a \ac{glmb} prior, explicit computation of the predicted measurement density is computationally challenging because it requires summation over possible object-to-measurement association hypotheses.
Instead, the discovered object \ac{phd} is readily obtained from the \ac{glmb} prior, from which $D_{\mathrm{v},k|k-1}$ is approximated as
\begin{align}
  D_{\mathrm{v},k|k-1}(\mathbf{z}; \mathcal{S})
  &\approx
  \int  D_{d, k|k-1}(\mathbf{x})
  p_{D,k}(\mathbf{x}; \mathcal{S})
  g_{k}(\mathbf{z}|\mathbf{x}) \mathrm{d} \mathbf{x} \nonumber \\
  \label{eq:PhdDiscoveredObjectMeasurement}
  &\qquad
  + \kappa_{c,k}(\mathbf{z})
\end{align}

\begin{figure}[htpb]
  \centering
  \def\svgwidth{\columnwidth}
\begingroup%
  \makeatletter%
  \providecommand\color[2][]{%
    \errmessage{(Inkscape) Color is used for the text in Inkscape, but the package 'color.sty' is not loaded}%
    \renewcommand\color[2][]{}%
  }%
  \providecommand\transparent[1]{%
    \errmessage{(Inkscape) Transparency is used (non-zero) for the text in Inkscape, but the package 'transparent.sty' is not loaded}%
    \renewcommand\transparent[1]{}%
  }%
  \providecommand\rotatebox[2]{#2}%
  \newcommand*\fsize{\dimexpr\f@size pt\relax}%
  \newcommand*\lineheight[1]{\fontsize{\fsize}{#1\fsize}\selectfont}%
  \ifx\svgwidth\undefined%
    \setlength{\unitlength}{491bp}%
    \ifx\svgscale\undefined%
      \relax%
    \else%
      \setlength{\unitlength}{\unitlength * \real{\svgscale}}%
    \fi%
  \else%
    \setlength{\unitlength}{\svgwidth}%
  \fi%
  \global\let\svgwidth\undefined%
  \global\let\svgscale\undefined%
  \makeatother%
  \begin{picture}(1,0.71283096)%
    \lineheight{1}%
    \setlength\tabcolsep{0pt}%
    \put(0,0){\includegraphics[width=\unitlength,page=1]{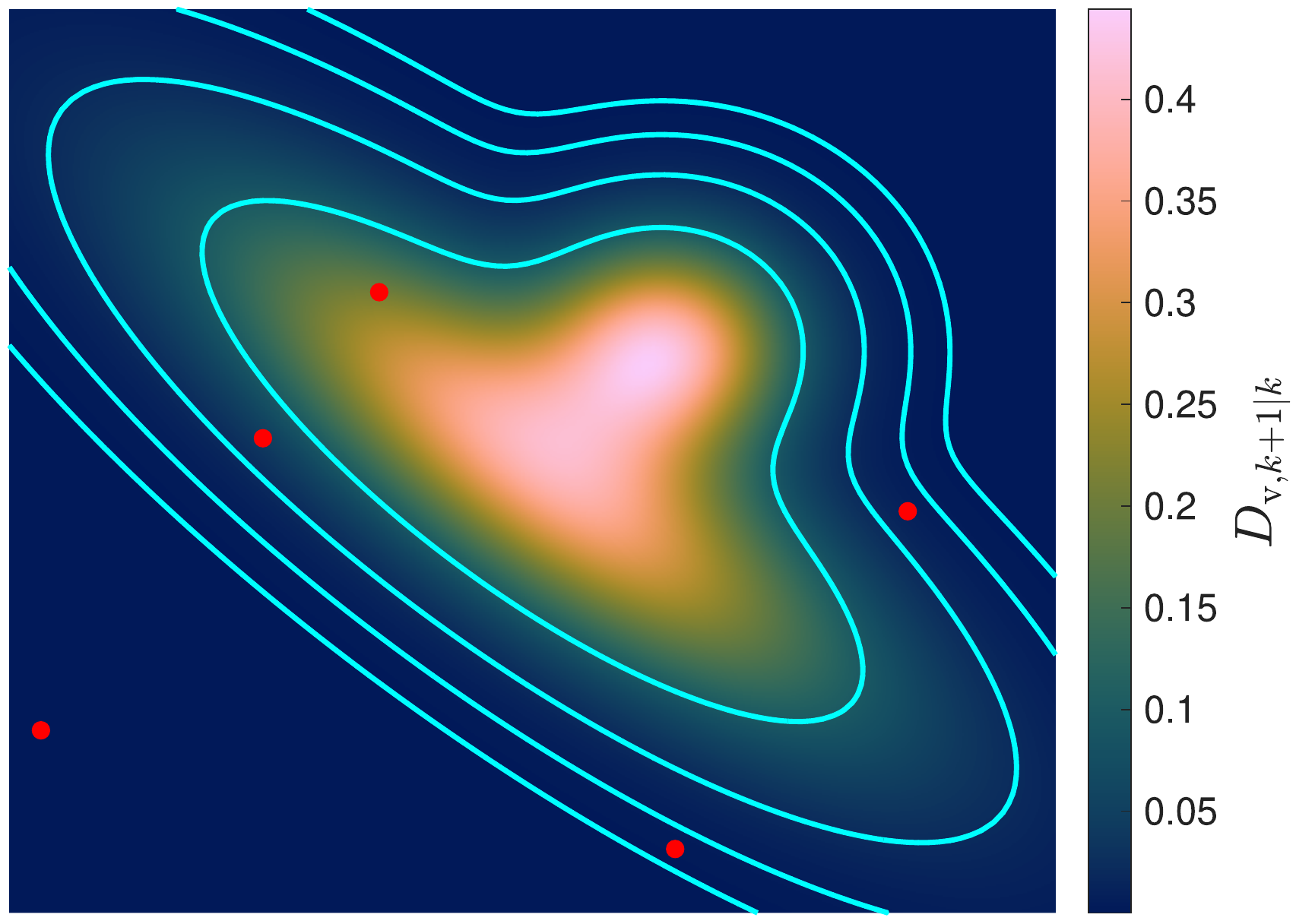}}%
    \put(0.25821099,0.06376638){\color[rgb]{1,1,1}\makebox(0,0)[lt]{\lineheight{1.25}\smash{\begin{tabular}[t]{l}$\overset{j}{\Omega}_{1}$\end{tabular}}}}%
    \put(0.36219067,0.03726208){\color[rgb]{1,1,1}\makebox(0,0)[lt]{\lineheight{1.25}\smash{\begin{tabular}[t]{l}$\overset{j}{\Omega}_{2}$\end{tabular}}}}%
    \put(0,0){\includegraphics[width=\unitlength,page=2]{FigHistogramQuadrature.pdf}}%
    \put(0.75175076,0.01771492){\color[rgb]{1,1,1}\makebox(0,0)[lt]{\lineheight{1.25}\smash{\begin{tabular}[t]{l}$\overset{j}{\Omega}_{3}$\end{tabular}}}}%
    \put(0.66589461,0.10752455){\color[rgb]{1,1,1}\makebox(0,0)[lt]{\lineheight{1.25}\smash{\begin{tabular}[t]{l}$\overset{j}{\Omega}_{4}$\end{tabular}}}}%
    \put(0.49236999,0.216515){\color[rgb]{1,1,1}\makebox(0,0)[lt]{\lineheight{1.25}\smash{\begin{tabular}[t]{l}$\overset{j}{\Omega}_{5}$\end{tabular}}}}%
  \end{picture}%
\endgroup%

  \caption{Example quadrature of the single-measurement conditional expected information gain, where representative measurements $\mathbf{z}_{j,i}$ are denoted by red dots and quadrature regions are outlined in cyan.}%
  \label{fig:HistogramQuadrature}
\end{figure}

Because no analytic solution of the integral in (\ref{eq:DiscoveredObjectConditionalReward}) exists, a numerical quadrature rule is employed.
In the proposed approach, a measurement cell is further tessellated into regions $\{\overset{j}{\Omega_{i}}\}_{i=1}^{R_{j}}\subset \overset{j}{\mathbb{Z}}$ based on the anticipated information value of measurements within each region, as illustrated in Fig.~\ref{fig:HistogramQuadrature}.
Then, given a representative measurement $\mathbf{z}_{j,i}$ for each region, the conditional information gain expectation is approximated as
\begin{align}
  \label{eq:HistogramQuadrature}
  \hat{\mathcal{R}}_{\mathrm{v}, k}^{d,j}(\overset{j}{\mathcal{S}})
  \approx
  \sum_{i=1}^{R_{j}} \mathcal{R}_{k}^{d}(\{\mathbf{z}_{j,i}\}; \overset{j}{\mathcal{S}}) p_{\mathrm{v}}^{j}(\mathbf{z}_{j,i}) A_{j,i}
\end{align}
where $A_{j,i}$ is the volume of region $\overset{j}{\Omega}_{i}$.
By this approach, the \ac{phd}-based \ac{kld} information gain function is only evaluated $R_{j}$ times.
Further details regarding the computation of the quadrature regions and representative measurement points are provided in Appendix~\ref{app:Quadrature}.

\subsection{Expected Information Gain of Undetected Objects}%
\label{sec:undiscovered_object_reward_expectation}
This subsection presents a new approach to efficiently model the undiscovered object distribution, which may be diffuse over a large region.
Although \acp{gm} and particle representations can be used to model undiscovered objects, they are highly inefficient at representing diffuse distributions.
Thus, in this paper, the position-marginal density of undiscovered objects is taken to be piecewise homogeneous with \ac{phd}
\begin{equation}
  \label{eq:UndiscoveredObjectPhd}
  D_{u,k|k-1}(\mathbf{s}) = \sum_{j=1}^{P} \frac{1_{\overset{j}{\mathbb{X}}_{s}} (\mathbf{s})}{
  A(\overset{j}{\mathbb{X}}_{s})}
  \cdot \lambda_{j,k|k-1}
\end{equation}
where $\lambda_{j,k|k-1}$ is the expected number of undiscovered objects in $\overset{j}{\mathbb{X}}_{s}$ at time step $k$ and $A(\overset{j}{\mathbb{X}}_{s})$ is the volume of cell $\overset{j}{\mathbb{X}}_{s}$.
For ease of exposition, the undiscovered object \ac{phd} is modeled using the same cell-decomposition employed in the \ac{cellmb} approximation.
Modeling undiscovered objects as a Poisson point process is one of the core ideas of the \ac{pmbm} filter, where discovered objects are modeled as a \ac{mbm} \ac{rfs}.
While discovered objects are modeled as a \ac{glmb} distribution in this work, the \ac{cellmb} \ac{swt} framework is amenable to any discovered object \ac{rfs} prior, including Poisson, \ac{iidc}, \ac{mb}, \ac{mbm}, \ac{lmb}, and \ac{glmb} distributions.

If $f_{k|k-1}(W_{k})$ is \ac{cellmb} with parameters $\{r_{\mathrm{w}}^{j}, p_{\mathrm{w}}^{j}\}_{j=1}^{P}$, then by Theorem \ref{thm:CellMbExpectation},
  \begin{align}
    \mathrm{E}[\mathcal{R}_{k}^{u}]
    =
    \sum_{j=1}^{P}
      \mathcal{R}_{k}^{u}(\emptyset; \overset{j}{\mathcal{S}}_{k})
      \left(1 - r_{\mathrm{w}}^{j}\right)%
      +
      \hat{\mathcal{R}}_{\mathrm{w},k}^{u,j}(\overset{j}{\mathcal{S}}_{k})
      \cdot
      r_{\mathrm{w}}^{j}
  \end{align}
  where
  \begin{align}
    \label{eq:UndiscoveredObjectConditionalReward}
    \hat{\mathcal{R}}_{\mathrm{w},k}^{u,j} (\overset{j}{\mathcal{S}})&\triangleq
    \int_{\overset{j}{\mathbb{Z}}} \mathcal{R}_{k}^{u}(\{\mathbf{z}\}; \overset{j}{\mathcal{S}}) p_{\mathrm{w}}^{j}(\mathbf{z}) \mathrm{d} \mathbf{z}\\
  r_{\mathrm{w}}^{j}(\mathcal{S}) &= \int 1_{\overset{j}{\mathbb{Z}}}(\mathbf{z}) D_{\mathrm{w},k|k-1}(\mathbf{z}; \mathcal{S}) \mathrm{d} \mathbf{z} \\
  &\approx
  \frac{\lambda_{j,k|k-1}}{A(\overset{j}{\mathbb{X}}_{s})} \int_{\overset{j}{\mathbb{X}}_{s}} p_{D}(\mathbf{s}; \mathcal{S}) \mathrm{d} \mathbf{s}\\
  p_{\mathrm{w}}^{j}(\mathbf{z}; \mathcal{S}) &= \frac{1}{r_{\mathrm{w}}^{j}} 1_{\overset{j}{\mathbb{Z}}}(\mathbf{z}) D_{\mathrm{w},k|k-1}(\mathbf{z}; \mathcal{S})\\
  D_{\mathrm{w},k|k-1}(\mathbf{z}; \mathcal{S})
  &=
  \int D_{u, k|k-1}(\mathbf{x})
  p_{D,k}(\mathbf{x}; \mathcal{S})
  g_{k}(\mathbf{z}|\mathbf{x}) \mathrm{d} \mathbf{x} \nonumber\\
  &\qquad
  + \kappa_{c,k}(\mathbf{z})
\end{align}

Under a piecewise homogeneous \ac{phd}, the undiscovered object information gain simplifies drastically if the measurement likelihood is independent of non-position states: i.e.~$g(\cdot|\mathbf{x})=g(\cdot| \mathbf{s})$.
Following (\ref{eq:KldPhdPseudolikelihood}),
\begin{align}
  \qquad\mathcal{R}_{k}^{u}&(W_{k};\, \mathcal{S}_{k}) \\
     &=
     \int_{\mathbb{X}_{s}} D_{u,k|k-1}(\mathbf{s})
     \{
       1 - L_{W_{k}}(\mathbf{s};\mathcal{S}_{k}) \nonumber \\
     &\quad+
     L_{W_{k}}(\mathbf{s};\mathcal{S}_{k})
     \log[L_{W_{k}}(\mathbf{s}; \mathcal{S}_{k})]
     \}   \mathrm{d} \mathbf{s} \nonumber
\end{align}
Given that at most one measurement may exist per cell, two cases need to be considered: the null measurement case and the singleton measurement case.
Letting $W_{k}=\emptyset$, and after some algebraic manipulation, the undiscovered object information gain for a null measurement can be written as
\begin{align}
  \label{eq:UndiscoveredObjectRewardNullMeasurement}
  \mathcal{R}_{k}^{u}(\emptyset; \mathcal{S}_{k})
  &=
  \sum_{j=1}^{P}
  \mathcal{R}_{k}^{u}(\emptyset; \overset{j}{\mathcal{S}}_{k})\\
  \label{eq:UndiscoveredObjectRewardNullMeasurementCell}
  \mathcal{R}_{k}^{u}(\emptyset; \overset{j}{\mathcal{S}}_{k})
  &=
  \lambda_{j,k|k-1} \cdot d_{j} \cdot (1 - \delta_{\emptyset}(\overset{j}{\mathcal{S}}_{k}))\\[1em]
  \label{eq:d_IntegralPdLog}
  d_{j}
  \triangleq
  \frac{1}{A(\overset{j}{\mathbb{X}}_{s})}
  &\int_{\overset{j}{\mathbb{X}}_{s}}
    p_{D}(\mathbf{s})  + (1 - p_{D}(\mathbf{s})) \log[1 - p_{D}(\mathbf{s})]
 \mathrm{d}\mathbf{s}
\end{align}
Furthermore, if the probability of detection is homogeneous within cells such that
\begin{equation}
  p_{D}(\mathbf{s}) = p_{D,j} \quad \forall \, \mathbf{s}\in \overset{j}{\mathbb{X}}_{s}
\end{equation}
then (\ref{eq:d_IntegralPdLog}) simplifies to
\begin{equation}
  d_{j} = p_{D,j} + (1-p_{D,j}) \log(1- p_{D,j})
\end{equation}

For the singleton measurement case, similar analytic simplifications of the conditional information gain (\ref{eq:UndiscoveredObjectConditionalReward}) are limited.
However, within a cell, the uniform position density of undiscovered objects is known \textit{a priori} up to an unknown factor $\lambda_{j,k|k-1}$.
Thus, the undiscovered object information gain can be pre-computed for efficiency and
\begin{equation}
  \label{eq:UndiscoveredRewardInterpolation}
  \hat{\mathcal{R}}_{\mathrm{w},k}^{u,j}(\overset{j}{\mathcal{S}}_{k}) \approx \bar{\mathcal{R}}_{\mathrm{w}}^{u,j}(\lambda_{j,k|k-1})
\end{equation}
where the function $\bar{\mathcal{R}}_{\mathrm{w}}^{u,j}(\lambda_{j,k|k-1})$ returns interpolated information gain values over $\lambda_{j,k|k-1}\in[0,1]$.

\textit{Remark:} In the special case that $p_D(\mathbf{s})=1$, the term $\mathcal{R}_{k}^{u}(\emptyset; \overset{j}{\mathcal{S}}_{k})$ is equivalent to the search objective term proposed in \cite{BostromRostSensorManagementPMBM21}.
The \ac{cellmb} approach differs in that perfect detection is not assumed, and that the information gained from detecting an undiscovered object, captured in the term $\hat{\mathcal{R}}_{\mathrm{w},k}^{u,j}$, is also considered.

\subsection{Field-of-View Optimization and Sensor Control}%
\label{sec:FieldOfRegard}
Prior to optimization of the \ac{fov}, the information gain associated with each cell in the \ac{for} is computed, as described in Algorithm~\ref{alg:FieldOfRegard}.
The \ac{for} cell information gains for discovered and undiscovered objects are stored as arrays $\{\mathcal{R}_{k}^{d}[j]\}_{j=1}^{P}$ and $\{\mathcal{R}_{k}^{u}[j]\}_{j=1}^{P}$, respectively.
Then, the optimal \ac{fov} is found as the one composed of the cells with the highest composite information gain, without reevaluating the information gain.
With this, the sensor control that produces the desired optimal \ac{fov} can be written as
\begin{equation}
  \label{eq:MaximizeRewardArray}
  \mathbf{u}_{k}^{*} = \argmax_{\mathbf{u} \in \mathbb{U}_{k}}
  \left\lbrace
    \sum_{j\in \mathbb{N}_{P}, \,  \overset{j}{\mathbb{X}}_{s} \subseteq \mathcal{S}_{k}(\mathbf{u})}
    \left(
      \mathcal{R}_{k}^{u}[j] + \mathcal{R}_{k}^{d}[j]
    \right)
  \right\rbrace
\end{equation}
where $\overset{j}{\mathcal{T}}_{k}\triangleq \mathcal{T}_{k} \cap \overset{j}{\mathbb{X}}_{s}$.

\begin{algorithm}[htbp]
  \caption{\Ac{for} Information Gain Pseudocode}
  \label{alg:FieldOfRegard}
  \hspace*{\algorithmicindent} \textbf{Input:}
  $\mathcal{T}_{k}$,
  $\mathring{f}_{k|k-1}(\mathring{X})$,
  $D_{u,k|k-1}(\mathbf{x})$ \\
\begin{algorithmic}
  \STATE Compute $D_{d,k|k-1}(\mathbf{x})$ from $\mathring{f}_{k|k}(\mathring{X})$ \hfill (\ref{eq:PhdDiscoveredObjectsFromGlmb})
  \STATE Compute $D_{\mathrm{v},k|k-1}(\mathbf{z};\mathcal{T}_{k})$ \hfill (\ref{eq:PhdDiscoveredObjectMeasurement})
  \FOR {$j=1,\dots,P$ for $j$ such that $\overset{j}{\mathbb{X}_{s}}\in \mathcal{T}_{k}$}
  \STATE $r_{\mathrm{v}}^{j} \gets \int 1_{\overset{j}{\mathbb{Z}}}(\mathbf{z}) D_{\mathrm{v},k|k-1}(\mathbf{z}; \mathcal{T}_{k}) \mathrm{d} \mathbf{z}$
  \STATE $r_{\mathrm{w}}^{j} \gets \int 1_{\overset{j}{\mathbb{Z}}}(\mathbf{z}) D_{\mathrm{w},k|k-1}(\mathbf{z}; \mathcal{T}_{k}) \mathrm{d} \mathbf{z}$
  \STATE Compute $\hat{\mathcal{R}}_{\mathrm{v}, k}^{d,j}(\overset{j}{\mathcal{T}})$ \hfill (\ref{eq:HistogramQuadrature})
  \STATE Compute $\hat{\mathcal{R}}_{\mathrm{w}, k}^{u,j}(\overset{j}{\mathcal{T}})$ \hfill (\ref{eq:UndiscoveredRewardInterpolation})
  \STATE $\mathcal{R}_{k}^{d}[j] \gets \mathcal{R}_{k}^{d}(\emptyset; \overset{j}{\mathcal{T}}_{k})(1-r_{\mathrm{v}}^{j}) + \hat{\mathcal{R}}_{\mathrm{v},k}^{d,j}(\overset{j}{\mathcal{T}}_{k})\cdot r_{\mathrm{v}}^{j}$ \hfill %
  \STATE $\mathcal{R}_{k}^{u}[j] \gets \mathcal{R}_{k}^{u}(\emptyset; \overset{j}{\mathcal{T}}_{k})(1-r_{\mathrm{w}}^{j}) + \hat{\mathcal{R}}_{\mathrm{w},k}^{u,j}(\overset{j}{\mathcal{T}}_{k})\cdot r_{\mathrm{w}}^{j}$ \hfill %
  \ENDFOR
\RETURN $(\mathcal{R}_{k}^{d}[j])_{j=1}^{P}, \, (\mathcal{R}_{k}^{u}[j])_{j=1}^{P}$
\end{algorithmic}
\end{algorithm}

\textit{Remark:} Explicit computation of the \ac{cellmb} single-object densities $p_{\mathrm{v}}^{j}$ and $p_{\mathrm{w}}^{j}$ is not required. Instead, these densities are implicitly computed when evaluating the conditional information gain expectations $\hat{\mathcal{R}}_{\mathrm{v},k}^{d,j}$ and $\hat{\mathcal{R}}_{\mathrm{w},k}^{u,j}$.

\subsection{Undiscovered Object Prediction and Update}%
\label{sec:undiscovered_object_prediction_and_update}
The prediction and update of the undiscovered object \ac{phd} is accomplished using the \ac{phd} filter, which we have discretized over cells.
The prediction step incorporates undiscovered object motion, birth, and death.
The undiscovered object distribution parameters are predicted and updated as
\begin{align}
  \label{eq:UndiscoveredObjectFilterPredict}
  \lambda_{j,k|k-1} &= \lambda_{B,j,k} + \sum_{i=1}^{P} p_{S,i,k} \cdot P_{j|i} \cdot \lambda_{i,k-1}\\
  \label{eq:UndiscoveredObjectFilterUpdate}
  \lambda_{j,k} &= \left[ 1 - p_{D,j}\cdot(1- \delta_{\emptyset}(\overset{j}{\mathcal{S}}_{k}))\right] \cdot \lambda_{j,k|k-1}
\end{align}
where $\lambda_{B,j,k}$ is the expected number of newborn objects in cell $j$, $p_{S,i,k}$ is the probability that an undiscovered object in cell $i$ survives, and $P_{j|i}$ is the probability that an undiscovered object moves to cell $j$ given that it exists in cell $i$.

\subsection{Discovered Object Tracking}%
\label{sec:discovered_object_tracking}
Discovered object tracking is performed using the data-driven \ac{glmb} filter. While a detailed description of the data-driven \ac{glmb} filter is beyond the scope of this paper, we highlight one important consideration involving the \ac{fov}-dependent nonlinear probability of detection.
The data-driven \ac{glmb} is implemented in \ac{gm} form, such that single-object densities are
\begin{equation}
  p^{(\xi)}(\mathbf{x}, \ell) = \sum_{i=1}^{J^{(\xi)}(\ell)} w_{i}^{(\xi)}(\ell) \mathcal{N}(\mathbf{x}; \boldsymbol{m}_{i}^{(\xi)}(\ell), \boldsymbol{P}_{i}^{(\xi)}(\ell))
\end{equation}
It is through the FoV-dependent $p_D$ that the filter probabilistically incorporates the knowledge of where objects were not seen.

\begin{figure}[h!]
\newlength\conefovgraphicheight
\setlength\conefovgraphicheight{4.8cm}
  \centering
  \subfloat[]{\includegraphics[height=\conefovgraphicheight]{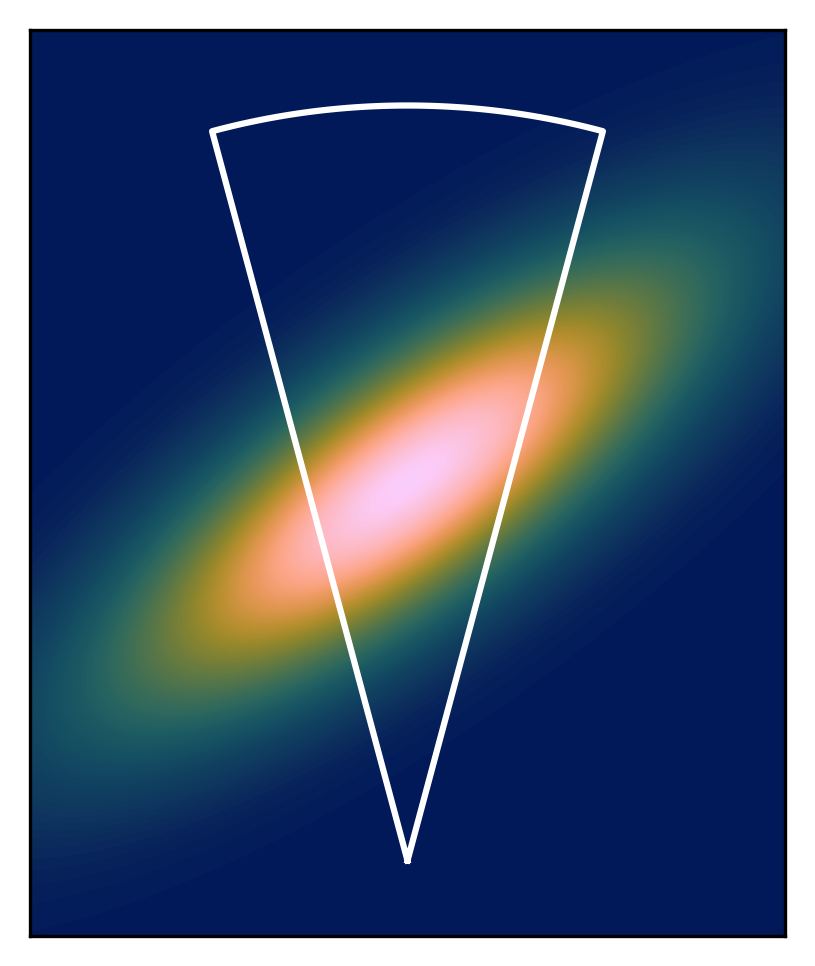}}\hfill
  \subfloat[]{\includegraphics[height=\conefovgraphicheight]{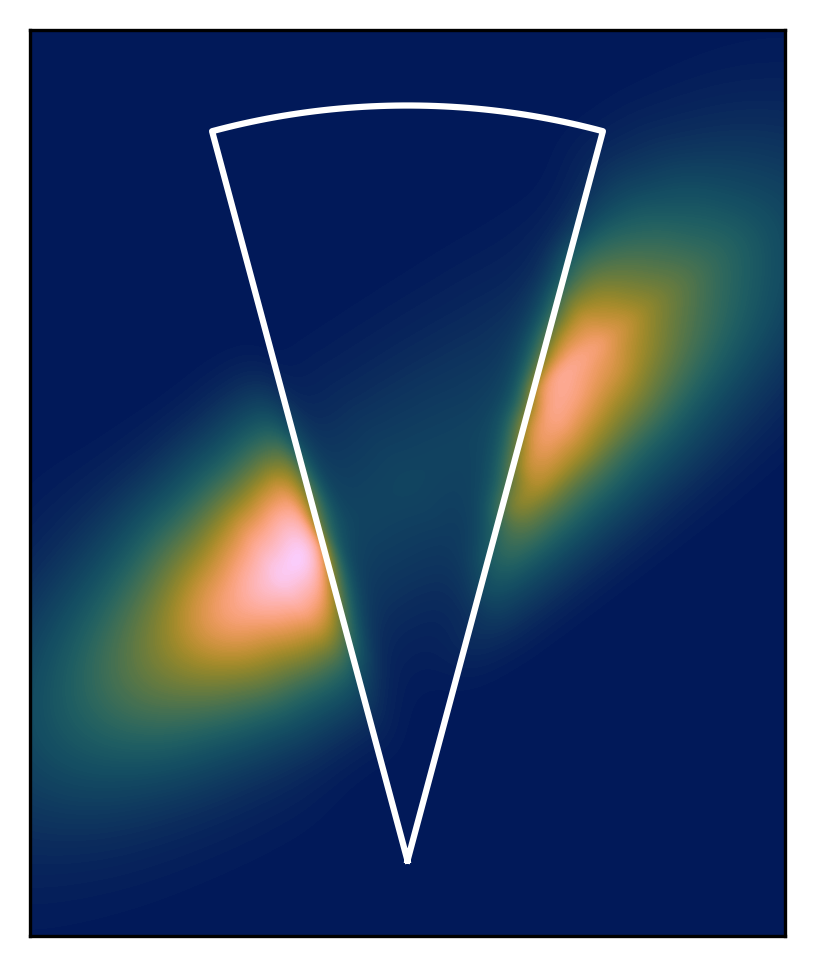}}%
  \caption{Prior object density and FoV (a), and posterior object density after recursive split and non-detection (b).}
  \label{fig:Example2d}
\end{figure}

In the filter, products of the form $p_{D}(\mathbf{x}; \mathcal{S})p^{(\xi)}(\mathbf{x})$ are expanded about the \ac{gm} component means in a zeroth-order Taylor expansion.
The accuracy of this approximation is dependent on the \ac{gm} resolution near the \ac{fov} boundaries.
Thus, a recursive splitting algorithm \cite{LeGrandRoleOfBoundedFovs20} employed that identifies and splits Gaussian components that overlap the \ac{fov} boundaries into several ``smaller'' Gaussians.
The resulting $J'^{(\xi)}(\ell)$ component mixture replaces the original density, enabling the accurate approximation
\begin{align}
  p_{D}(\mathbf{x}; \mathcal{S}) & p^{(\xi)}(\mathbf{x}, \ell) \nonumber \\
  \approx \sum_{i=1}^{J'^{(\xi,\ell)}} &w_{i}^{(\xi,\ell)} p_{D}(\boldsymbol{m}_{i}^{(\xi,\ell)};\mathcal{S}) \mathcal{N}(\mathbf{x}; \boldsymbol{m}_{i}^{(\xi,\ell)}, \boldsymbol{P}_{i}^{(\xi,\ell)})
\end{align}
An example is provided in Fig.~\ref{fig:Example2d}, wherein the prior density is split prior to a Bayes update, allowing for the accurate incorporation of negative information from a non-detection.

\subsection{Numerical Implementation}%
\label{sec:algorithm_overview}
This subsection summarizes the \ac{swt} algorithmic implementation.
At each step $k$,  a time-update (\ref{eq:LabeledBayesFilterPredict}), (\ref{eq:UndiscoveredObjectFilterPredict}) of the previous posterior densities $\mathring{f}(\mathring{X}_{k-1}|Z_{0:k-1})$ and $D_{u,k-1}$ yields predicted prior densities for the time of the next decision.
The \ac{for} is constructed from admissible control actions as shown in (\ref{eq:FieldOfRegardConstruction}), and the expected information gain for each cell within the \ac{for} is computed.
The candidate \ac{fov} that contains the maximizing sum of cell information gains is found, and the corresponding control (\ref{eq:MaximizeRewardArray}) which yields that \ac{fov} is applied.
The sensor collects a new multi-object measurement which is processed in the data-driven GLMB filter to update the multi-object density, giving the posterior density in (\ref{eq:DataDrivenLabeledBayesFilterUpdate}).
The algorithm is summarized in Algorithm~\ref{alg:SearchDetectTrack}.

\begin{algorithm*}[htbp]
  \caption{SWT Sensor Control Pseudocode}
  \label{alg:SearchDetectTrack}
  \hspace*{\algorithmicindent} \textbf{Input:} $\mathring{f}_{0}(\mathring{X})$, $D_{u,0}(\mathbf{x})$ \\
\begin{algorithmic}
  \FOR {$k=1,\hdots, K$}
  \STATE $\mathring{f}_{k|k-1}(\mathring{X}), \, D_{u,k|k-1}(\mathbf{x}) \gets \texttt{filter\_prediction}(\mathring{f}_{k-1}(\mathring{X}),\, D_{u,k-1}(\mathbf{x}))$ \hfill (\ref{eq:LabeledBayesFilterPredict}), (\ref{eq:UndiscoveredObjectFilterPredict})
  \STATE $(\mathcal{R}_{k}^{d}[j])_{j=1}^{P}, \, (\mathcal{R}_{k}^{u}[j])_{j=1}^{P} \gets \texttt{FoR\_information\_gain}(\mathcal{T}_{k},
  \mathring{f}_{k|k-1}(\mathring{X}),
  D_{u,k|k-1}(\mathbf{x}))$ \hfill Alg.\,\ref{alg:FieldOfRegard}
\STATE $\mathbf{u}^{*}_{k} \gets \texttt{maximize\_expected\_reward}((\mathcal{R}_{k}^{d}[j])_{j=1}^{P}, \, (\mathcal{R}_{k}^{u}[j])_{j=1}^{P}) $\hfill(\ref{eq:MaximizeRewardArray})
  \STATE $\mathcal{S}_{k}(\mathbf{u}^{*}_{k}) \gets$ apply sensor control
  \STATE $Z_{k} \gets$ obtain measurement
  \STATE $\mathring{f}_{k|k-1}(\mathring{X}) \gets \texttt{split\_for\_FoV}(\mathring{f}_{k|k-1}(\mathring{X}), \, \mathcal{S}_{k})$ \hfill \cite{LeGrandRoleOfBoundedFovs20}
  \STATE $\mathring{f}_{k|k}(\mathring{X}), \, D_{u,k|k} \gets \texttt{filter\_update}(\mathring{f}_{k|k-1}(\mathring{X}), D_{u,k|k-1}(\mathbf{x}), \, Z_{k}, \, \mathcal{S}_{k})$
  \hfill (\ref{eq:DataDrivenLabeledBayesFilterUpdate}), (\ref{eq:UndiscoveredObjectFilterUpdate})
  \ENDFOR
\end{algorithmic}
\end{algorithm*}

\section{Application to Remote Multi-Vehicle SWT}%
\label{sec:application_to_vehicle_tracking}

\begin{figure}[htpb]
  \centering
  \includegraphics[width=0.97\linewidth]{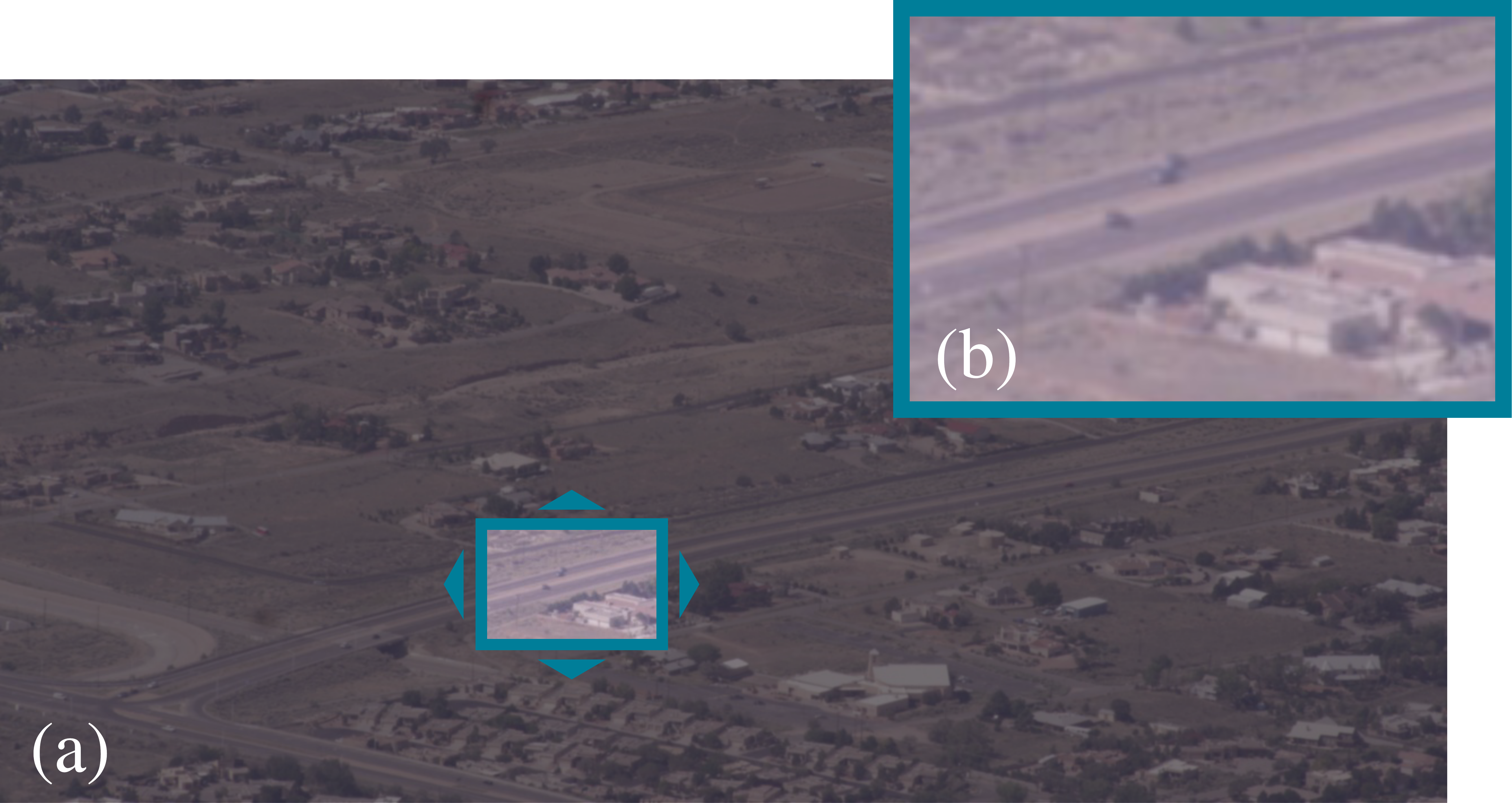}
  \caption{Example video frame (a) from Albuquerque dataset, artificially windowed to emulate smaller, movable FoV, which is enlarged in (b) to show detail.}%
  \label{fig:SandiaCrestFrameWindowed}
\end{figure}

The \ac{cellmb} \ac{swt} framework is demonstrated in two distinct vehicle tracking problems using real video data.
The first experiment, hereon referred to as the ``Albuquerque'' experiment, is based on a video recorded by a fixed camera pointed at a remote location where multiple mobile ground vehicles are observed.
The second ``Sydney'' experiment involves tracking multiple mobile maritime surface vehicles using real satellite video\footnote{Video publicly available at https://mall.charmingglobe.com.} taken of Sydney, Australia from the Chinese low Earth-orbiting satellite, Jilin-1.
In both experiments, real-time  \ac{fov} controlled motion is simulated by windowing the data over a small fraction of the available frame, as illustrated in Fig.~\ref{fig:SandiaCrestFrameWindowed}.
These datasets present significant tracking challenges, including jitter-induced noise and clutter, unknown measurement origin, merged detections from closely-spaced vehicles, and most significantly, temporal sparsity of detections.

\subsection{Vehicle Dynamics}%
\label{sec:vehicle_dynamics}
Vehicle dynamics are modeled directly in the image frame. While vehicle dynamics are more naturally expressed in the terrestrial frame, the cameras' precise location and orientation are unknown. Thus, the transformation between image and terrestrial coordinates could not be readily established
for the Albuquerque experiment.
The Sydney video is georegistered such that world coordinate motion maps directly to scaled image coordinate motion.

The object state is modeled as
\begin{gather}
  \mathbf{x}_{k} = [\mathbf{s}_{k}^{T} \quad \boldsymbol{\zeta}_{k}^{T}]^{T}\\[1em]
  \mathbf{s}_{k} = [\xi_{k} \quad \eta_{k} ]^{T} \,, \qquad \boldsymbol{\zeta}_{k}=[\dot{\xi}_{k} \quad \dot{\eta}_{k} \quad \Omega_{k}]^{T}
\end{gather}
where $\xi_{k}$ and $\eta_{k}$ are the horizontal and vertical coordinates, respectively, of the vehicle position with respect to the full-frame origin, $\dot{\xi}_{k}$ and $\dot{\eta}_{k}$ are the corresponding rates, and $\Omega_{k}$ is the vehicle turn rate.

\begin{figure}[htpb]
  \centering
  \ifx\undefined\usetikz
    \includegraphics[width=\linewidth]{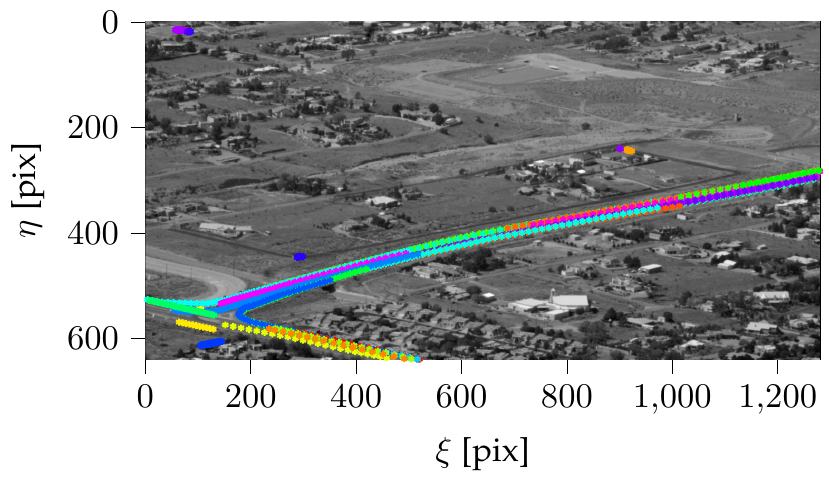}
  \else
    \import{figures/}{FrameWithGroundTruthTrajs.tikz}
  \fi
  \caption{True trajectories of moving objects with an example image as frame as background.}%
  \label{fig:FigFrameWithGroundTruthTrajs}
\end{figure}

Vehicle motion is modeled using the nearly coordinated turn model with directional process noise \cite{BarshalomTrackingDataAssociation11, YeddanapudiImmAirTrafficDirectionalNoise97} as
\begin{equation}
  \mathbf{x}_{k+1} = \mathbf{f}_{k}(\mathbf{x}_{k}) + \boldsymbol{\Gamma}_{k} \boldsymbol{\nu}_{k}(\mathbf{s}_{k})
\end{equation}
where $\mathbf{f}_{k}$ is defined in \cite[Ch.\ 11]{BarshalomEstimationNavigation01} and
\begin{equation}
  \boldsymbol{\Gamma}_{k} =
  \begin{bmatrix}
    \frac{1}{2} (\Delta t)^{2} \mathbf{I}_{2\times 2} & \boldsymbol{0}_{2 \times 1} \\
    (\Delta t) \mathbf{I}_{2 \times 2} & \boldsymbol{0}_{2\times 1} \\
    \boldsymbol{0}_{1 \times 2} & \Delta t
  \end{bmatrix}
\end{equation}
where $\Delta t = 1\, [\textrm{sec}]$ is the discrete time step interval, $\mathbf{I}_{n \times n}$ denotes the $n \times n$ identity matrix, and $\boldsymbol{0}_{m\times n}$ denotes the $m\times n$ matrix whose elements are zero.
The covariance of the process noise is
\begin{gather}
  \mathrm{E}[\boldsymbol{\nu}_{k} \boldsymbol{\nu}_{k}^{T}]=
  \mathbf{Q}_{k}(\mathbf{s})
  =
  \begin{bmatrix}
    \mathbf{D}^T(\mathbf{s}) \mathbf{Q}_{d} \mathbf{D}(\mathbf{s}) & 0 \\
    \boldsymbol{0}_{1 \times 2} & \sigma_{\Omega,\texttt{<ABQ,SYD>}}^{2}
  \end{bmatrix}\\
  \mathbf{Q}_{d} = \begin{bmatrix}
    \sigma_{t,\texttt{<ABQ,SYD>}}^{2} & 0 \\
    0 & \sigma_{n,\texttt{<ABQ,SYD>}}^{2}
  \end{bmatrix} \\
  \mathbf{D}(\mathbf{s})=
  \begin{bmatrix}
    \cos \Psi(\mathbf{s}) & \sin \Psi(\mathbf{s}) \\
    -\sin \Psi(\mathbf{s}) & \cos \Psi(\mathbf{s})
  \end{bmatrix}
\end{gather}

where $\sigma_{\Omega,\texttt{ABQ}}= 180\, [\textrm{arcmin}/\textrm{sec}]$ and $\sigma_{\Omega,\texttt{SYD}}= 30\, [\textrm{arcmin}/\textrm{sec}]$ are the turn rate process noise standard deviations, $\sigma_{t,\texttt{ABQ}}=5 \, [\textrm{pixel}/\textrm{sec}^2]$
and $\sigma_{n,\texttt{ABQ}}=0.01 \, [\textrm{pixel}/\textrm{sec}^2]$ are the standard deviation of process noise tangential and normal to the road, respectively, and $\Psi(\mathbf{s})$ is the angle of the road segment nearest $\mathbf{s}$, measured from the horizontal axis to the tangent direction.
Information-driven sensor control efficacy fundamentally depends on the accuracy of motion prediction and the rate of uncertainty growth in the absence of observations.
Evaluation of multiple candidate motion models revealed that the nearly coordinated turn model with directional process noise offered better motion prediction and more precise uncertainty growth over the simpler constant velocity models.
Road geometry is not applicable to the Sydney experiment, and thus an isotropic linear process noise is used with $\sigma_{t,\texttt{SYD}}=\sigma_{n,\texttt{SYD}}=1 \, [\textrm{pixel}/\textrm{sec}^2]$.
The true trajectories of all moving objects in the Albuquerque experiment are shown in Fig.~\ref{fig:FigFrameWithGroundTruthTrajs}.

\subsection{Sensor and Scene Model}%
\label{sec:SceneSenor}
Object detections are generated from raw frame data using normalized difference change detection \cite{SimonsonChangeDetectionJitter09} and fast approximate power iteration subspace tracking \cite{BadeauFapi05} for temporal background estimation.
The single-object measurement function is linear-Gaussian with corresponding likelihood
\begin{gather}
  g(\mathbf{z}|\mathbf{x}) = \mathcal{N}(\mathbf{z}; \, \mathbf{H} \mathbf{x}, \, \mathbf{R})\,,\\
  \mathbf{H} =
  \begin{bmatrix}
    \mathbf{I}_{2\times 2}  & \boldsymbol{0}_{2\times 3}
  \end{bmatrix} \,,
  \qquad
  \mathbf{R} = \sigma_{z,\texttt{<ABQ,SYD>}}^{2} \cdot \mathbf{I}_{2\times 2} \,
\end{gather}
where $\sigma_{z,\texttt{ABQ}}^{2}=9 \, [\textrm{pixel}^{2}]$ and~$\sigma_{z,\texttt{SYD}}^{2}=100 \, [\textrm{pixel}^{2}]$.
The Albuquerque experiment sensor \ac{fov} is a rectangular region that is $240$ pixels wide and $160$ pixels tall.
Rectangular \ac{fov} geometry is also emulated in the Sydney experiment, where the \ac{fov} dimensions are $1024$ pixels wide and $640$ pixels tall.

Moving objects within the \ac{fov} are assumed to be detectable with probability $p_{D,k}(\mathbf{s}_{k})=0.9$.

The mean false alarm rates are assumed to be five and thirty false detections per scene frame in the Albuquerque and Sydney experiments, respectively.

\begin{figure}[thpb]
  \centering
  \includegraphics[width=0.95\linewidth]{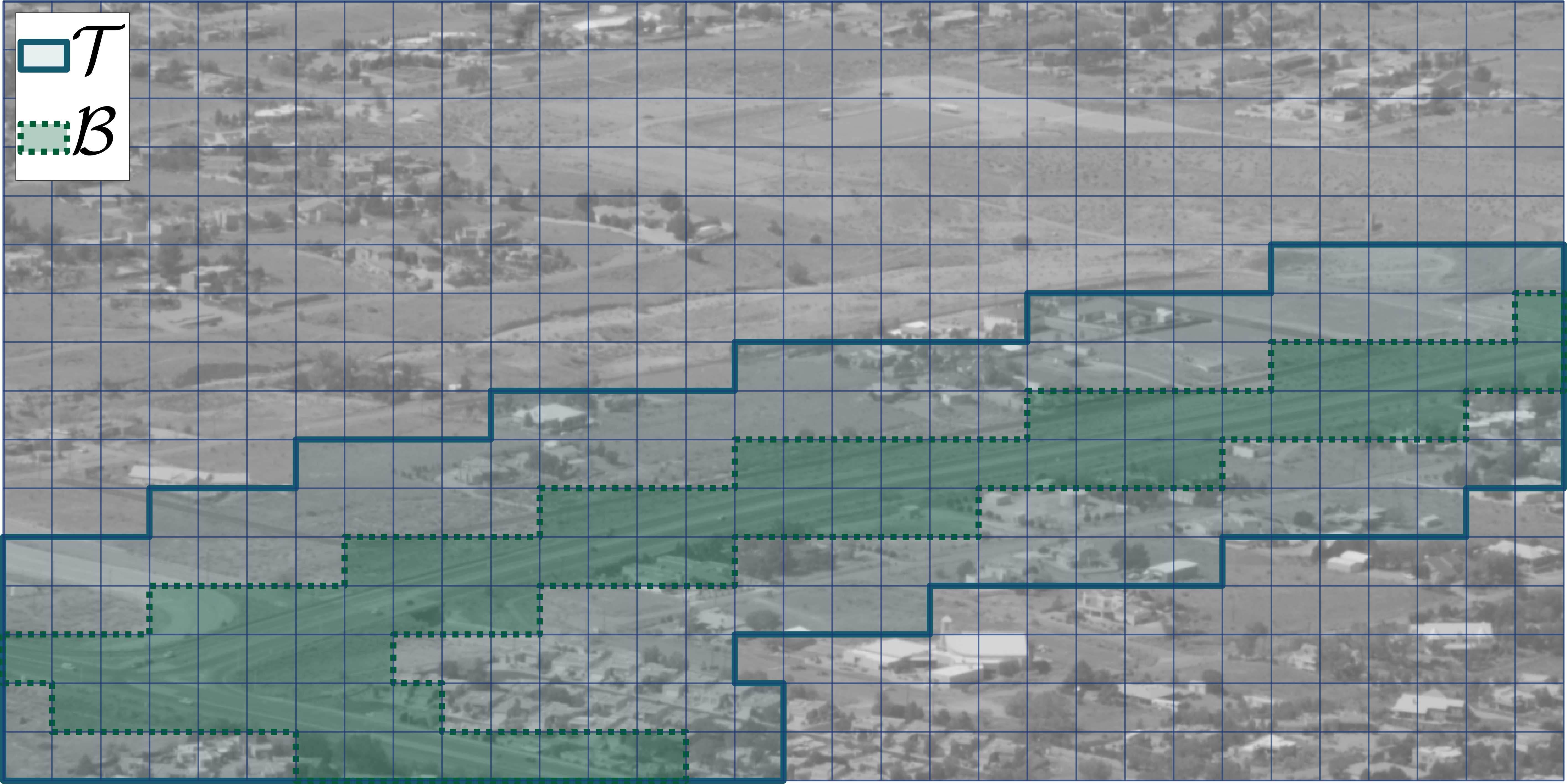}
  \caption{Albuquerque field-of-regard, $\mathcal{T}$, and primary road region $\mathcal{B}$, with example image frame as background.}%
  \label{fig:FrameWithCellRegions}
\end{figure}

\begin{figure}[thpb]
  \centering
  \includegraphics[width=0.95\linewidth]{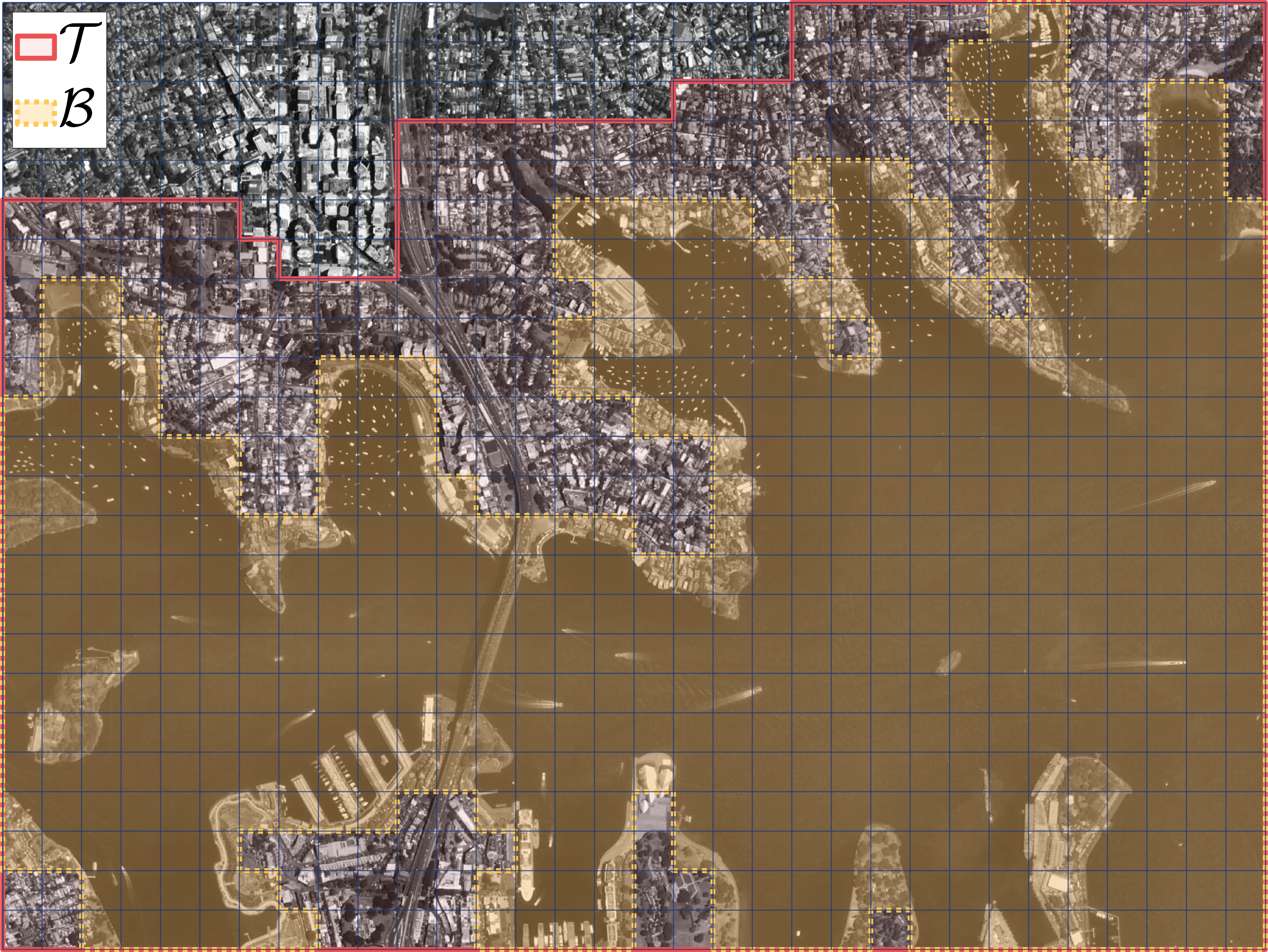}
  \caption{Sydney field-of-regard, $\mathcal{T}$, and water region $\mathcal{B}$, with example image frame as background.}%
  \label{fig:FrameWithCellRegionsBoats}
\end{figure}

The Albuquerque and Sydney scenes are tessellated by $16 \times 32$  and $24 \times 32$ grids, respectively, of uniformly sized rectangular cells as shown in Figs.~\ref{fig:FrameWithCellRegions} and~\ref{fig:FrameWithCellRegionsBoats}.
Within the Albuquerque scene, an \ac{roi} is specified which contains the scene's two primary roads and is denoted by $\mathcal{T}$ due to its equivalence to the \ac{for} for this problem.
Within the Albuquerque \ac{roi}, cells containing road pixels comprise the set $\mathcal{B}$, which is used to establish an initial uniform distribution of undiscovered objects.
In the Sydney experiment, the \ac{roi} is defined as the main water region, including the piers and wharves.
Thus, following the assumptions established in Section~\ref{sec:undiscovered_object_reward_expectation}, the initial undiscovered object position marginal \ac{phd} is characterized by (\ref{eq:UndiscoveredObjectPhd}) with

\begin{equation}
    \lambda_{j,0} = \begin{cases}
    \lambda_{\texttt{<ABQ},\texttt{SYD>}} & \overset{j}{\mathbb{X}} \subseteq \mathcal{B} \\
    0 & \textrm{otherwise}
  \end{cases}
\end{equation}
where $\lambda_{\texttt{ABQ}}=0.137$ and $\lambda_{\texttt{SYD}}=0.0593$ correspond to initial estimates of ten and thirty undiscovered objects in the scene, respectively.

\subsection{Experiment Results}%
\label{sec:experiment_results}
The Albuquerque and Sydney experiments consist of $60$ and $64$ time steps, respectively.
To emulate a pan/tilt camera from the wider available frame data, the \ac{fov} is assumed to be able to be moved to any location within the scene in a single time step. This is a reasonable assumption as these adjustments would be less than a degree.

\begin{figure}[tphb]
  \newlength{\framewidth}\setlength{\framewidth}{\columnwidth}
  \centering
  \includegraphics[trim=0 0 0 5cm, clip, width=\framewidth]{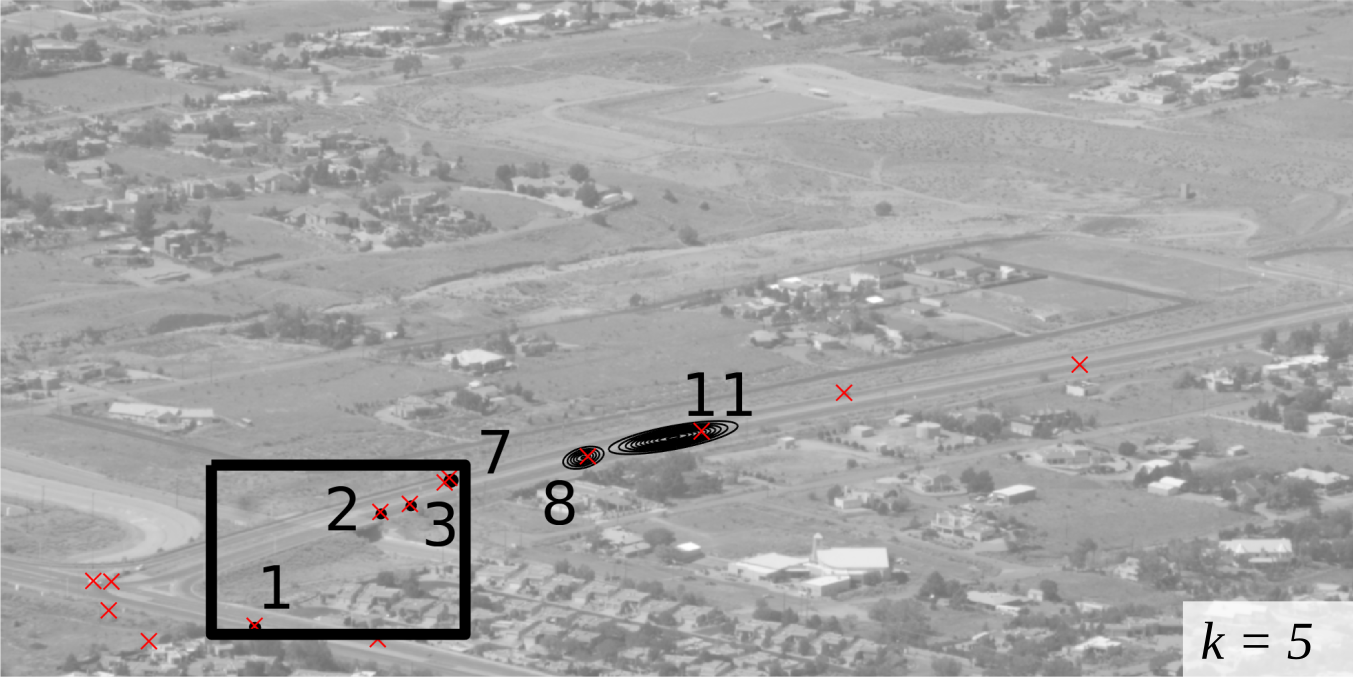}\\[1mm]
    \includegraphics[trim=0 0 0 5cm, clip, width=\framewidth]{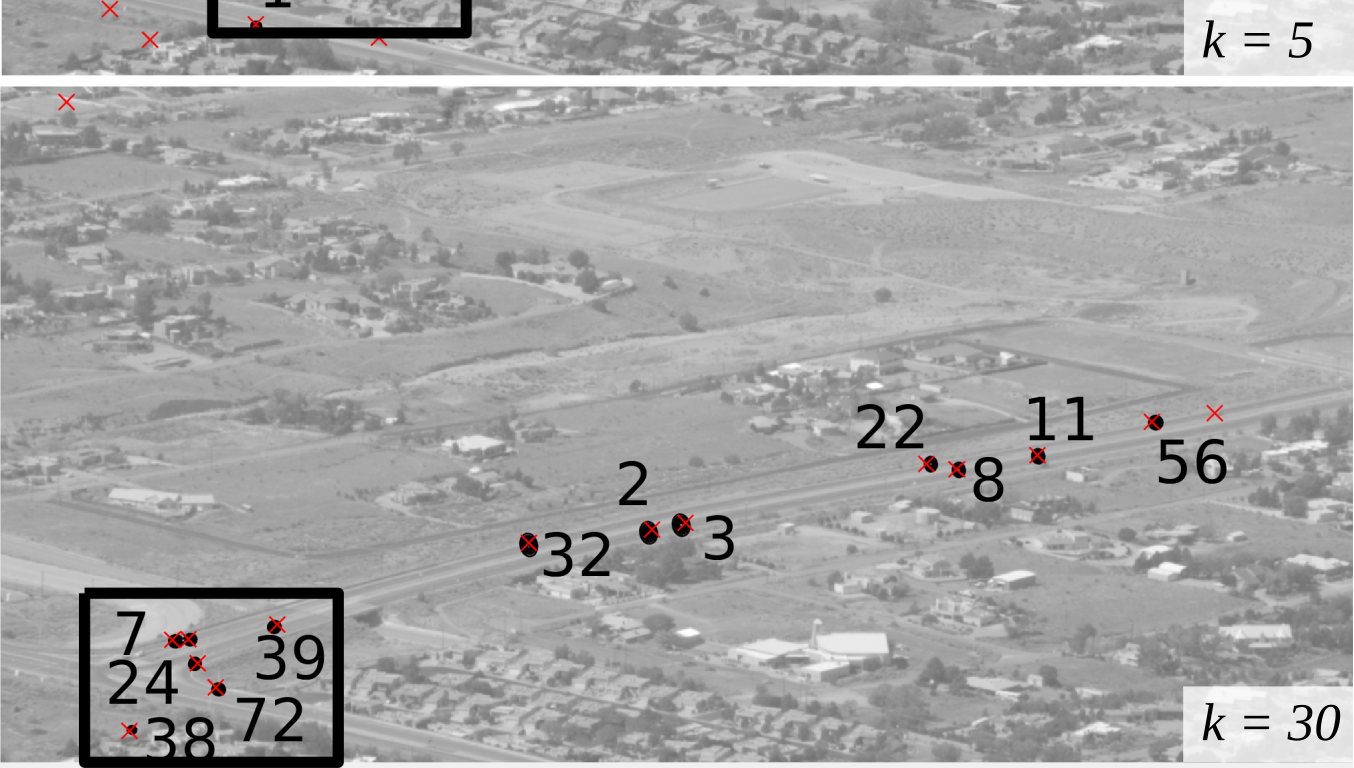}\\[1mm]
    \includegraphics[trim=0 0 0 5cm, clip, width=\framewidth]{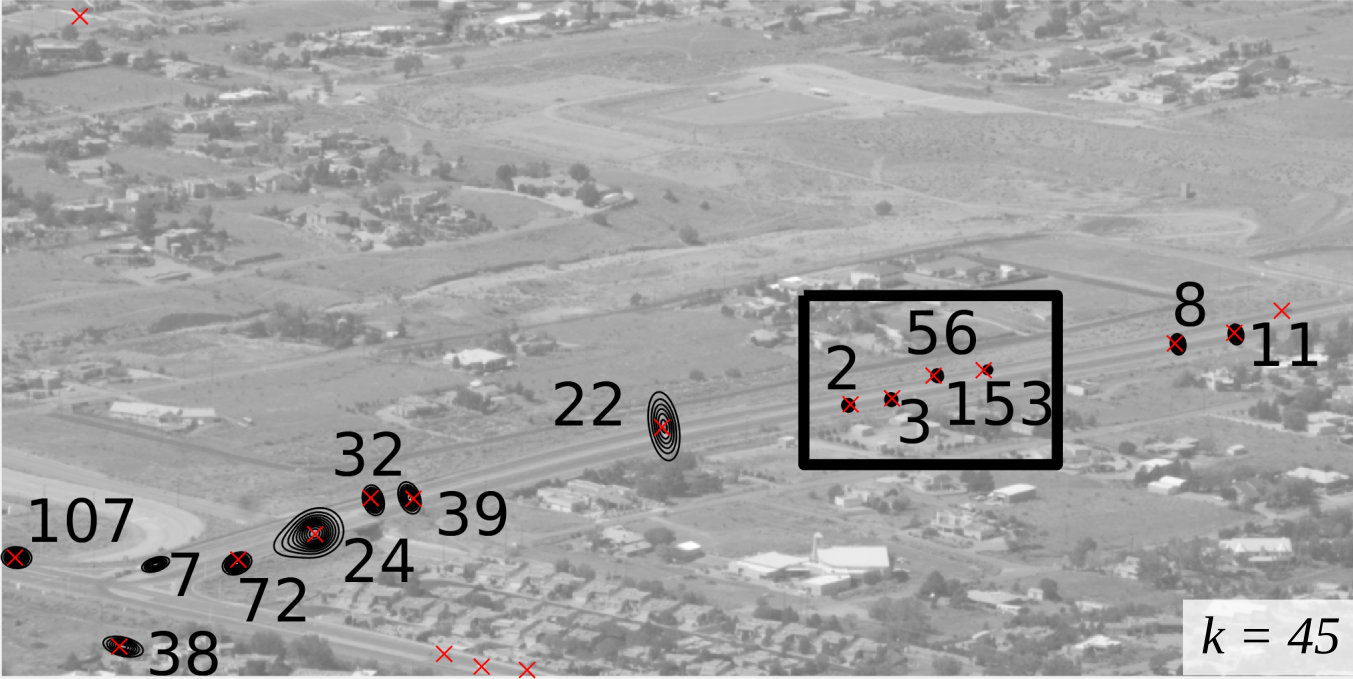}
  \caption{\ac{fov} position and tracker estimates in the form of single-object density contours for objects with probabilities of existence greater than $0.5$, shown at select time steps for the Albuquerque experiment.}%
  \label{fig:FramesWithLmbOverlaid}
\end{figure}

\begin{figure}[tphb]
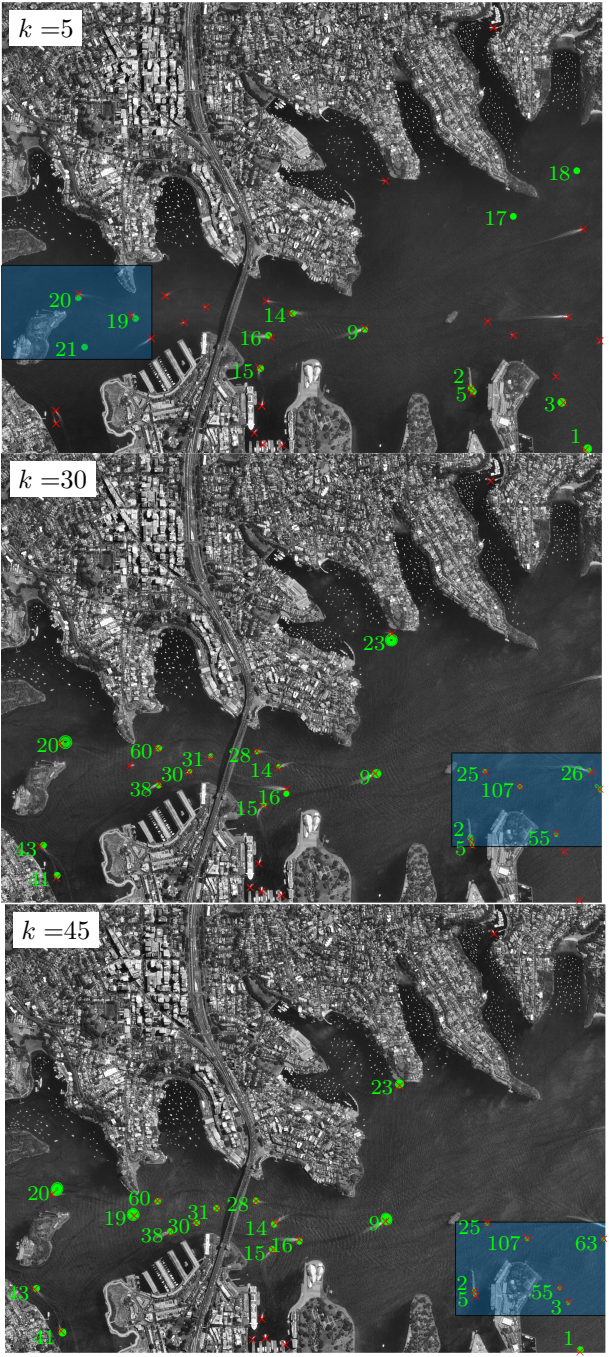

\newlength\lmbwithframewidth
\setlength\lmbwithframewidth{4.5cm}
\setlength{\figurewidth}{0.9\columnwidth}
  \centering
  \ifx\undefined\usetikz
  \newcounter{num}
  \setcounter{num}{0}
  \makeatletter
    \@for\k:={0,1,2}%
    \do{%
      \@for\s:={0}%
      \do{%
        \includegraphics{figures/FigFrameWithLmbBoats\arabic{num}}%
        \stepcounter{num}%
      }\\%
  }
  \makeatother
  \else
  \foreach \datestr in {2022_05_21_21_43}{
    \foreach \k in {05,30,45}{
      \newcommand{\labelfile}{figures/\datestr_k0\k_plot_labels.csv}%
      \newcommand{\imagefile}{figures/\datestr_k0\k_frame_with_lmb_no_labels.png}%
      \import{figures/}{FrameWithLmbFunction.tex}}
    }
  \fi
  \caption{\acs{fov} position and tracker estimates in the form of single-object density contours for objects with probabilities of existence greater than $0.5$, shown at select time steps for the Sydney experiment.}
  \label{fig:FramesWithLmbBoats}
\end{figure}

Some key frames of the Albuquerque and Sydney experiments are shown in Fig.~\ref{fig:FramesWithLmbOverlaid} and~\ref{fig:FramesWithLmbBoats}, respectively.
In the early time steps, the \ac{fov} motion is dominated by the undiscovered object component of the information gain.
As more objects are discovered and tracked, the observed actions demonstrate a balance of revisiting established tracks to reduce state uncertainty and exploring new areas where undiscovered objects may exist.

\begin{figure}[htpb]
  \centering
  \setlength{\figureheight}{1.4\columnwidth}
  \setlength{\figurewidth}{0.9\columnwidth}
  \ifx\undefined\usetikz
    \includegraphics[width=\linewidth]{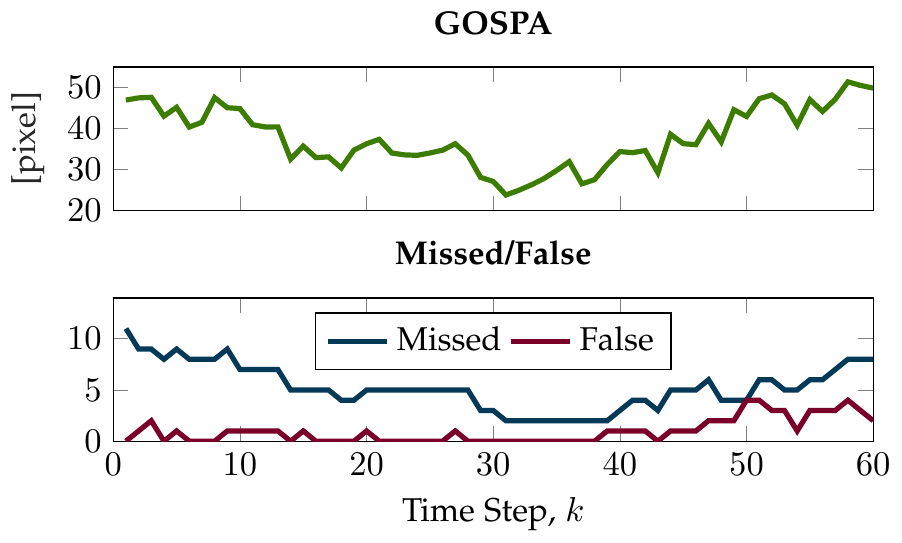}
  \else
    \import{figures/}{GospaSandiaCrest.tikz}
  \fi
  \caption{Albuquerque dataset SWT performance in terms of GOSPA metric and component errors over time using cutoff distance $c=20\, [\textrm{pixel}]$, order $p=2$, and $\alpha=2$.}%
  \label{fig:GospaSandiaCrest}
\end{figure}

\begin{figure}[htpb]
  \centering
  \setlength{\figureheight}{1.4\columnwidth}
  \setlength{\figurewidth}{0.85\columnwidth}
  \ifx\undefined\usetikz
    \includegraphics[width=\linewidth]{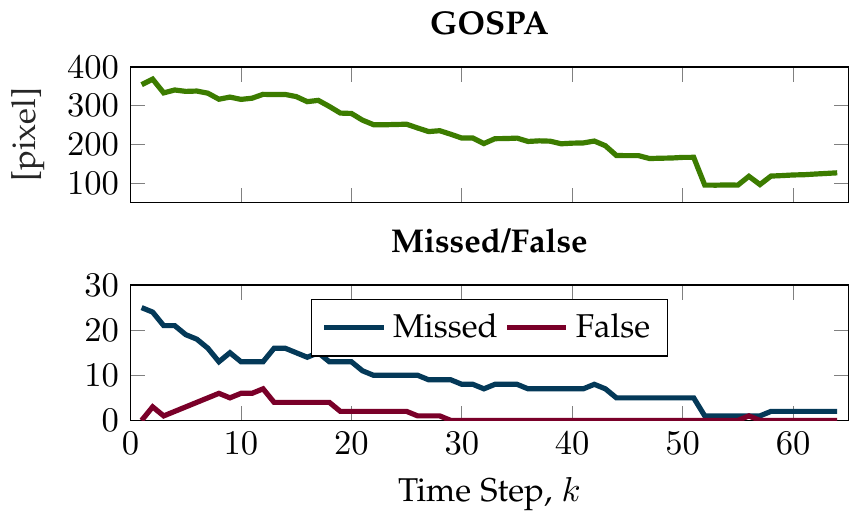}
  \else
    \import{figures/}{GospaBoats.tikz}
  \fi
  \caption{Sydney dataset SWT performance in terms of GOSPA metric and component errors over time using cutoff distance $c=100\, [\textrm{pixel}]$, order $p=2$, and $\alpha=2$.}%
  \label{fig:GospaBoats}
\end{figure}

Because the overall objective of the \ac{swt} sensor control is to reduce multi-object tracking uncertainty,  \ac{swt} sensor control performance is most naturally quantified by the resulting multi-object tracking accuracy, as measured using the \ac{gospa} metric \cite{RahmathullahGospa17}.
For the metric parameters selected in this work, the \ac{gospa} metric is equal to the sum of localization errors for properly tracked objects and penalties for missed and false tracks.
The \ac{gospa} metric and the number of false and missed objects are shown over time for the Albuquerque and Sydney experiments in Figs.~\ref{fig:GospaSandiaCrest} and~\ref{fig:GospaBoats}, respectively.
The \ac{cellmb} \ac{swt} sensor control effectively balances the competing objectives of new object discovery and maintenance of established tracks, as illustrated by the decline in missed objects and consistently low number of false tracks.
An increase in \ac{gospa} is observed in the final time steps of the Albuquerque experiment, which is caused by a sharp uptick in new object appearances.

The average \ac{gospa} over the experiment is compared with the \ac{pims}-based information driven control and random \ac{fov} motion in Table~\ref{tab:Gospa}.
The \ac{cellmb} sensor control achieves significant improvement with respect to other methods in the number of missed and false tracks, as well as the overall \ac{gospa} metric, which encompasses cardinality errors and localization errors.
\newcommand*\rot{\rotatebox{90}}
\begin{table}[h]
\caption{GOSPA performance on Albuquerque and Sydney experiments, averaged over experiment duration, with percentage improvement over baseline random control shown parenthetically.}
  \label{tab:Gospa}
  \begin{center}
  {%
    \begin{tabular}{cl|*{3}{r}}
    \toprule[2pt]
    & Algorithm      & GOSPA $\, [\textrm{pixel}]$ & Missed & False\\
    \cmidrule{1-5}
    & Cell-MB        & 37.84 \,(56\%) & 5.27 (168\%)   & 0.97 (158\%)\\
    & PIMS           & 47.46 (24\%) & 9.95 (42\%)  & 0.90 (178\%)\\
    \rot{\rlap{~\texttt{ABQ}}}& Random         & 59.07 (N\!/\!A) & 14.10 (N\!/\!A) & 2.50 (N\!/\!A)\\
    \cmidrule[1pt]{2-5}
    & Cell-MB	&  225.00 (41\%)  &  9.03 (85\%) 	 & 1.41 (97\%)\\
    & PIMS    &	 267.44	 (19\%) &  11.98 (39\%)	 & 2.31 (20\%)\\
    \rot{\rlap{~\texttt{SYD}}}& Random	&  318.00	(N\!/\!A)   &  16.67 (N\!/\!A)	 & 2.78 (N\!/\!A)%
  \end{tabular}
  }
  \end{center}
\end{table}
While the \ac{pims} approach exhibits degraded performance in these applications, it should still be considered as a viable method when using an information gain function that is not cell-additive.

\section{Conclusion}
This paper presents a novel \acf{cellmb} approximation that enables the tractable higher-order approximation of the expectation of set functions that are additive over disjoint measurable subsets.
The \ac{cellmb} approximation is useful for approximating the expectation of computationally-expensive set functions, such as information-theoretic reward functions employed in sensor control applications.
The approach is developed in the context of information-driven sensor control in which the objective is to discover and track an unknown time-varying number of non-cooperative objects with minimal estimation error.
The problem is formulated as a \acl{pomdp} with a new Kullback-Leibler divergence-based information gain that incorporates both discovered and undiscovered object information gain.

In demonstrations using real terrestrial and satellite sensor data, the \acl{swt} sensor control is used to manipulate the sensor fields-of-view to discover and track multiple moving ground vehicles and boats from an aerial vantage point.

\bibliographystyle{IEEEtran}
\bibliography{FISST_Keith_refs,silvia_refs_all}
\clearpage
\newpage
\appendices
\section{KLD minimization}
\label{app:KldMinimizingCellMb}
\begin{IEEEproof}
  The minimization of the \ac{kld} between $f$ and $\bar{f}$ can be equivalently expressed as the maximization problem
\begin{equation}
  \label{eq:KldMinimization}
  \max_{\{r^{j}, p^{j}\}_{j=1}^{P}}
  \left[
    \int f(Y) \log(\bar{f}(Y)) \delta Y
  \right]
\end{equation}
Equation~(\ref{eq:CellMbDensity}) can be equivalently written as
\begin{align}
  \bar{f}(Y) &=
  \Delta (Y, \mathbb{Y})
  \prod_{j=1}^{P} \left(1 - r^{j}\right)
  \left(
    \prod_{i=1}^{n}
    \sum_{j=1}^{P}
    \frac{r^{j} p^{j}(\mathbf{y}_{i})}{1 - r^{j}}
  \right)
  \label{eq:CellMbDensityProductForm}
\end{align}
where it is noted that for each $i$ in the rightmost product, the sum has only one non-zero term at $j=j'$ where $\overset{j'}{\mathbb{Y}}\ni \mathbf{y}_{i}$.
Thus, (\ref{eq:CellMbDensityProductForm}) can be factored as
\begin{align}
  \label{eq:CellMbDensityProductFormSplit}
  \bar{f}(Y) &=
  \Delta (Y, \mathbb{Y})
  \prod_{j=1}^{P} \left(1 - r^{j}\right)
  \left(
    \prod_{i=1}^{n}
    \sum_{j=1}^{P}
    \frac{1_{\overset{j}{\mathbb{Y}}}(\mathbf{y}_{i})}{1 - r^{j}}
  \right) \\
  &\qquad \cdot
  \left(
    \prod_{i=1}^{n}
    \sum_{j=1}^{P}
    r^{j} p^{j_{n}}(\mathbf{y}_{i})
  \right)
  \nonumber
\end{align}
According to (\ref{eq:PhdOfMbRfsDensity}), the rightmost sum of (\ref{eq:CellMbDensityProductFormSplit}) is equal to the \ac{phd} of $\bar{f}(Y)$, such that
\begin{align}
  \bar{f}(Y) &=
  \Delta (Y, \mathbb{Y})
  \prod_{j=1}^{P} \left(1 - r^{j}\right)
  \left(
    \prod_{i=1}^{n}
    \sum_{j=1}^{P}
    \frac{1_{\overset{j}{\mathbb{Y}}}(\mathbf{y}_{i})}{1 - r^{j}}
  \right)
  \left(
    \prod_{i=1}^{n}
    \bar{D}(\mathbf{y}_{i})
  \right)
  \label{eq:CellMbDensityPhd}
\end{align}
Now, taking the logarithm of (\ref{eq:CellMbDensityPhd}),
\begin{align}
  \log(\bar{f}(Y))
  &=
  \log(
  \Delta (Y, \mathbb{Y})
  )
  +
  \log
  \prod_{j=1}^{P} \left(1 - r^{j}\right)\\
  &\qquad
  +
  \log
  \left(
    \prod_{i=1}^{n}
    \sum_{j=1}^{P}
    \frac{1_{\overset{j}{\mathbb{Y}}}(\mathbf{y}_{i})}{1 - r^{j}}
  \right)
  +
  \log
  \left(
    \prod_{i=1}^{n}
    \bar{D}(\mathbf{y}_{i})
  \right) \nonumber \\
  &=
  \log(
  \Delta (Y, \mathbb{Y})
  )
  +
  \sum_{j=1}^{P}
  \log
  \left( 1 - r^{j}\right)\\
  \label{eq:LogOfFbar}
  &\qquad
  +
  \sum_{i=1}^{n}
  \log
  \left(
    \sum_{j=1}^{P}
    \frac{1_{\overset{j}{\mathbb{Y}}}(\mathbf{y}_{i})}{1 - r^{j}}
  \right)
  +
  \sum_{i=1}^{n}
  \log
  \left(
    \bar{D}(\mathbf{y}_{i})
  \right) \nonumber
\end{align}
The third term in (\ref{eq:LogOfFbar}) can be modified by recognizing that the inner sum has only one non-zero term, and thus
\begin{align}
  \sum_{i=1}^{n}
  \log
  \left(
    \sum_{j=1}^{P}
    \frac{1_{\overset{j}{\mathbb{Y}}}(\mathbf{y}_{i})}{1 - r^{j}}
  \right)
  &=
  \sum_{i=1}^{n}
  \log
  \left(
    \frac{1}{1 - \sum_{j=1}^{P} 1_{\overset{j}{\mathbb{Y}}}(\mathbf{y}_{i})r^{j}}
  \right)\\
  &=
  -\sum_{i=1}^{n}
  \log
  \left(
    1 - \sum_{j=1}^{P} 1_{\overset{j}{\mathbb{Y}}}(\mathbf{y}_{i})r^{j}
  \right)
\end{align}
which, by substitution of into (\ref{eq:LogOfFbar}), yields
\begin{align}
  \label{eq:LogCellMbDensityPhd}
  \log(\bar{f}(Y))
  &=
  \log(
  \Delta (Y, \mathbb{Y})
  )
  +
  \sum_{j=1}^{P}
  \log
  \left( 1 - r^{j}\right)\\
  &\qquad
  -
  \sum_{i=1}^{n}
  \log
  \left(
    1 - \sum_{j=1}^{P} 1_{\overset{j}{\mathbb{Y}}}(\mathbf{y}_{i})r^{j}
  \right)
  \nonumber \\
  &\qquad
  +
  \sum_{i=1}^{n}
  \log
  \left(
    \bar{D}(\mathbf{y}_{i})
  \right)
  \nonumber
\end{align}
Now taking the set integral of the product
\begin{align}
  \int & f(Y) \log(\bar{f}(Y)) \delta Y \nonumber \\
  &=
  \int f(Y)
  \log(
  \Delta (Y, \mathbb{Y})
  )
  \delta Y
  +
  \int
  \sum_{j=1}^{P}
  f(Y)
  \log
  \left( 1 - r^{j}\right)
  \delta Y \nonumber\\
  &\qquad
  -
  \int
  \sum_{i=1}^{n}
  f(Y)
  \log
  \left(
    1 - \sum_{j=1}^{P} 1_{\overset{j}{\mathbb{Y}}}(\mathbf{y}_{i})r^{j}
  \right)
  \delta Y \nonumber \\
  &\qquad
  +
  \int
  \sum_{i=1}^{n}
  f(Y)
  \log
  \left(
    \bar{D}(\mathbf{y}_{i})
  \right)
  \delta Y \\[1em]
  &=
  \int
  \sum_{j=1}^{P}
  f(Y)
  \log
  \left( 1 - r^{j}\right)
  \delta Y \nonumber\\
  &\qquad
  -
  \int
  \sum_{i=1}^{n}
  f(Y)
  \log
  \left(
    1 - \sum_{j=1}^{P} 1_{\overset{j}{\mathbb{Y}}}(\mathbf{y}_{i})r^{j}
  \right)
  \delta Y \nonumber \\
  \label{eq:IntfLogf_1}
  &\qquad
  +
  \int
  \sum_{i=1}^{n}
  f(Y)
  \log
  \left(
    \bar{D}(\mathbf{y}_{i})
  \right)
  \delta Y
\end{align}
where in the last equation, the first term vanished due to the observation that $f(Y)=0$ everywhere that $\Delta(Y,\mathbb{Y})=0$ and by application of the identity $\lim_{x\to 0} x \log x = 0$.
By applying Proposition 2a of \cite{MahlerFirstOrderMoments03}, which states that
\begin{equation}
  \int f(Y) \sum_{i=1}^{n} h(\mathbf{y}_{i}) \delta Y = \int D(\mathbf{y}) h(\mathbf{y}) \mathrm{d} \mathbf{y}
\end{equation}
equation (\ref{eq:IntfLogf_1}) can be rewritten in terms of the \ac{phd} as
\begin{align}
  \int & f(Y) \log(\bar{f}(Y)) \delta Y \nonumber \\
  &=
  \sum_{j=1}^{P}
  \log
  \left( 1 - r^{j}\right)
  -
  \int
  D(\mathbf{y})
  \log
  \left(
    1 - \sum_{j=1}^{P} 1_{\overset{j}{\mathbb{Y}}}(\mathbf{y})r^{j}
  \right)
  \mathrm{d} \mathbf{y} \nonumber \\
  &\qquad
  +
  \int
  D(\mathbf{y})
  \log
  \left(
    \bar{D}(\mathbf{y})
  \right)
  \mathrm{d} \mathbf{y} \\[1em]
  &=
  \sum_{j=1}^{P}
  \log
  \left( 1 - r^{j}\right)
  -
  \sum_{j=1}^{P}
  \int
  1_{\overset{j}{\mathbb{Y}}}(\mathbf{y})
  D(\mathbf{y})
  \log
  \left(
    1 - r^{j}
  \right)
  \mathrm{d} \mathbf{y} \nonumber \\
  &\qquad
  +
  \int
  D(\mathbf{y})
  \log
  \left(
    \bar{D}(\mathbf{y})
  \right)
  \mathrm{d} \mathbf{y} \\[1em]
  &=
  \sum_{j=1}^{P}
  \log
  \left( 1 - r^{j}\right)
  \left( 1 - \int 1_{\overset{j}{\mathbb{Y}}}(\mathbf{y}) D(\mathbf{y})\right) \nonumber\\
  &\qquad
  +
  \int
  D(\mathbf{y})
  \log
  \left(
    \bar{D}(\mathbf{y})
  \right)
  \mathrm{d} \mathbf{y}
  \label{eq:IntfLogf}
\end{align}
Equation~(\ref{eq:IntfLogf}) consists of the sum of two terms and is maximized when both terms are simultaneously maximized.
The first term is maximized by
\begin{align*}
  r^{j} = \int 1_{\overset{j}{\mathbb{Y}}}(\mathbf{y}) D(\mathbf{y}) \mathrm{d} \mathbf{y}
\end{align*}
and the second term is maximized when $\bar{D}(\mathbf{y})= D(\mathbf{y})$.
By (\ref{eq:PhdOfMbRfsDensity}),
\begin{align}
  \bar{D}(\mathbf{y})
  &=
  \sum_{j=1}^{P} r^{j} p^{j}(\mathbf{y})
  =
  \sum_{j=1}^{P}
  1_{\overset{j}{\mathbb{Y}}}(\mathbf{y}) D(\mathbf{y})
\end{align}
By equating like-terms,
\begin{align}
  p^{j}(\mathbf{y}) &= \frac{1}{r^{j}} 1_{\overset{j}{\mathbb{Y}}}(\mathbf{y}) D(\mathbf{y})
\end{align}
completing the proof.
\end{IEEEproof}

\section{Cell-MB Expectation}
\label{app:CellMbExpectationProof}

Let $\mathbf{z}_{1:n}\triangleq \mathbf{z}_{1}, \hdots, \mathbf{z}_{n}$ and $\mathrm{d}\mathbf{z}_{1:n}\triangleq\mathrm{d}\mathbf{z}_{1} \cdots \mathrm{d}\mathbf{z}_{n}$.
Substitution of the cell-MB density (\ref{eq:CellMbDensity}) and cell-additive information gain (\ref{eq:RewardAdditiveOverZ}) into (\ref{eq:RewardExpectation})
gives
\begin{align}
  \mathrm{E}[\mathcal{R}]
  &=
  \left[\frac{r^{(\cdot)}}{1 - r^{(\cdot)}}\right]^{\mathbb{N}_{P}}
  \left(
    \mathcal{R}(\emptyset; \mathcal{S})
    +
    \sum_{n=1}^{P}
    \frac{1}{n!}
    \psi(n; \mathcal{S})
  \right)
    \label{eq:ExpectedRewardProof0}
\end{align}
where
\begin{align*}
  \psi(n; \mathcal{S})
  &\triangleq
  \int
  \Delta(\{\mathbf{z}_{1:n}\}, \mathbb{Z})\nonumber \\
  &\quad\cdot
    \left[
      \sum_{j=1}^{P}
      \mathcal{R} (\{\mathbf{z}_{1:n}\} \cap \overset{j}{\mathbb{Z}}; \overset{j}{\mathcal{S}})
    \right]
    \left[
      \sum_{j=1}^{P}
      \frac{r^{j}p^{j}(\mathbf{z}_{(\cdot)})}{1 - r^{j}}
    \right]^{\mathbb{N}_{n}}
    \!\!\!
    \mathrm{d} \mathbf{z}_{1:n}
\end{align*}
We wish to simplify the set integral into combinations of vector integrals $\int \cdot \mathrm{d} \mathbf{z}$ such that the expected information gain is computationally feasible.
Integrals on $\mathbb{Z}$ are equivalent to
\begin{equation}
  \int h(\mathbf{z}) \mathrm{d} \mathbf{z} =
  \int_{\overset{1}{\mathbb{Z}}} h\big(\overset{1}{\mathbf{z}}\big) \mathrm{d} \overset{1}{\mathbf{z}}
  + \cdots +
  \int_{\overset{P}{\mathbb{Z}}} h\big(\overset{P}{\mathbf{z}}\big) \mathrm{d} \overset{P}{\mathbf{z}}
\end{equation}
where $\overset{j}{\mathbf{z}}\in \overset{j}{\mathbb{Z}}$, as shown in \cite[Eqn.~3.50]{MahlerAdvancesStatisticalMultitargetFusion14}.
First, note that the rightmost sum
\begin{equation}
  \sum_{j=1}^{P}
  \frac{r^{j}p^{j}(\mathbf{z}_{i})}{1 - r^{j}}
\end{equation}
has only one non-zero term: namely, when $j=j'$ where $\mathbf{z}_i \in \overset{j'}{\mathbb{Z}}$.
Then, the integral can be written as a sum of integrals, each wherein $\mathbf{z}_{1}$ is integrated over a different subset $\overset{i_{1}}{\mathbb{Z}} \subseteq \mathbb{Z}, \quad i_{1} \in \mathbb{N}_{P}$ as follows:
\begin{align}
    \psi(n;\mathcal{S})
    &=
    \sum_{i_{1}=1}^{P}
    \int \Delta(\{\mathbf{z}_{2:n}\}, \overline{\mathbb{Z}}(i_{1})) \nonumber \\
    &\quad\cdot
    \left[
      \mathcal{R} (\{\overset{i_{1}}{\mathbf{z}}\}; \overset{i_{1}}{\mathcal{S}})
      +
      \sum_{j=1,j\neq i_{1}}^{P}
      \mathcal{R} (\{\mathbf{z}_{2:n}\} \cap \overset{j}{\mathbb{Z}}; \overset{j}{\mathcal{S}})
    \right] \nonumber \\
    &\cdot
    \left(\frac{r^{i_{1}}p^{i_{1}}(\overset{i_{1}}{\mathbf{z}})}{1 - r^{i_{1}}}\right)
    \left[
      \sum_{j=1, j\neq i_{1}}^{P}
      \frac{r^{j}p^{j}(\mathbf{z}_{(\cdot)})}{1 - r^{j}}
    \right]^{\mathbb{N}_{n} \setminus \mathbb{N}_{1}}
    \mathrm{d} \overset{i_{1}}{\mathbf{z}}\,  \mathrm{d} \mathbf{z}_{2:n}
\end{align}
  where
  \begin{equation}
  \overline{\mathbb{Z}}(i_{1},...,i_{n}) \triangleq
  \mathbb{Z} \setminus (\overset{i_{1}}{\mathbb{Z}} \uplus \cdots \uplus \overset{i_{n}}{\mathbb{Z}})
  \end{equation}
Repeating the same procedure for $\mathbf{z}_{2},...,\mathbf{z}_{n}$
\begin{align}
  &\psi(n;\mathcal{S}) \nonumber \\
     &=
    \sum_{i_{1}=1}^{P}
    \sum_{i_{2}=1, i_{2}\neq i_{1}}^{P}
    \int \Delta(\{\mathbf{z}_{3:n}\}, \overline{\mathbb{Z}}(i_{1}, i_{2})) \nonumber\\
     &\quad \cdot
    \left[
      \mathcal{R} (\{\overset{i_{1}}{\mathbf{z}}\}; \overset{i_{1}}{\mathcal{S}})
      +
      \mathcal{R} (\{\overset{i_{2}}{\mathbf{z}}\}; \overset{i_{2}}{\mathcal{S}})
      +
      \sum_{\mathclap{j=1,j\notin \{i_{1}, i_{2}\}}}^{P}
      \mathcal{R} (\{\mathbf{z}_{3:n}\} \cap \overset{j}{\mathbb{Z}}; \overset{j}{\mathcal{S}})
    \right] \nonumber \\
     &\quad
    \cdot
    \left(\frac{r^{i_{(\cdot)}}p^{i_{(\cdot)}}(\overset{i_{(\cdot)}}{\mathbf{z}})}{1 - r^{i_{(\cdot)}}}\right)^{\mathbb{N}_{2}}
    \left[
      \sum_{j=1, j \notin\{ i_{1}, i_{2}\}}^{P}
      \frac{r^{j}p^{j}(\mathbf{z}_{(\cdot)})}{1 - r^{j}}
    \right]^{\mathbb{N}_{n}\setminus\mathbb{N}_{2}} \nonumber \\
     &\qquad \cdot
    \mathrm{d} \overset{i_{1}}{\mathbf{z}}\,
    \mathrm{d} \overset{i_{2}}{\mathbf{z}}\,
    \mathrm{d} \mathbf{z}_{3:n} \\[1em]
    &=
    \sum_{1\leq i_{1} \neq \cdots \neq i_{n} \leq P}
    \int
    \bigg[
      \mathcal{R} (\{\overset{i_{1}}{\mathbf{z}}\}; \overset{i_{1}}{\mathcal{S}})
      + \cdots +
      \mathcal{R} (\{\overset{i_{n}}{\mathbf{z}}\}; \overset{i_{n}}{\mathcal{S}})
      +
    \nonumber \\
    &\qquad
      \mathcal{R} (\emptyset; \overline{\mathcal{S}}(i_{1}, \hdots i_{n}))
    \bigg]
    \left[\frac{r^{i_{(\cdot)}}p^{i_{(\cdot)}}(\overset{i_{(\cdot)}}{\mathbf{z}})}{1 - r^{i_{(\cdot)}}}\right]^{\mathbb{N}_{n}}
    \mathrm{d} \overset{i_{1}}{\mathbf{z}} \cdots
    \mathrm{d} \overset{i_{n}}{\mathbf{z}}
\end{align}
Moving the existence probability terms outide the integral and exploiting symmetry over order permuations of $(i_{1},\hdots, i_{n})$,
\begin{align}
    &\psi(n;\mathcal{S}) \nonumber \\
    &=
    n! \hspace{-0.4cm}
    \sum_{1\leq i_{1} < \cdots < \neq i_{n} \leq P}
    \left[\frac{r^{i_{(\cdot)}}}{1 - r^{i_{(\cdot)}}}\right]^{\mathbb{N}_{n}} \nonumber \\
    &\quad
    \int
    \left[
      \mathcal{R} (\{\overset{i_{1}}{\mathbf{z}}\}; \overset{i_{1}}{\mathcal{S}})
      + \cdots +
      \mathcal{R} (\{\overset{i_{n}}{\mathbf{z}}\}; \overset{i_{n}}{\mathcal{S}})
      +
      \mathcal{R} (\emptyset; \overline{\mathcal{S}}(i_{1}, \hdots i_{n}))
    \right] \nonumber \\
    &\qquad
    \cdot
    \left[
      p^{i_{(\cdot)}}(\overset{i_{(\cdot)}}{\mathbf{z}})
    \right]^{\mathbb{N}_{n}}
    \mathrm{d} \overset{i_{1}}{\mathbf{z}} \cdots
    \mathrm{d} \overset{i_{n}}{\mathbf{z}} \\[1em]
    &=
    n!
    \sum_{1\leq i_{1} < \cdots < \neq i_{n} \leq P}
    \left[\frac{r^{i_{(\cdot)}}}{1 - r^{i_{(\cdot)}}}\right]^{\mathbb{N}_{n}} \nonumber \\
    &\quad \cdot
    \left[
      \mathcal{R} (\emptyset; \overline{\mathcal{S}}(i_{1}, \hdots i_{n}))
      +
      \sum_{\ell=1}^{n}
      \int
      \mathcal{R} (\{\overset{i_{\ell}}{\mathbf{z}}\}; \overset{i_{\ell}}{\mathcal{S}}) p^{i_{\ell}} (\overset{i_{\ell}}{\mathbf{z}})
    \mathrm{d} \overset{i_{\ell}}{\mathbf{z}}
    \right]
    \label{eq:ExpectedRewardSumSingleIntegrals}
\end{align}
where the last line is obtained by using the pdf property $\int p^{i_{\ell}}(\overset{i_{\ell}}{\mathbf{z}}) \mathrm{d} \overset{i_{\ell}}{\mathbf{z}}=1$.
Substitution of (\ref{eq:ExpectedRewardSumSingleIntegrals}) into (\ref{eq:ExpectedRewardProof0}) gives
  \begin{align}
    \mathrm{E}[\mathcal{R}] &=
    \left[1 - r^{(\cdot)}\right]^{\mathbb{N}_{P}}
      \mathcal{R}(\emptyset; \mathcal{S})
      +
    \Bigg[
      \sum_{n=1}^{P}
      \sum_{1\leq i_{1} < \cdots < i_{n} \leq P}
      \left[r^{i_{(\cdot)}}\right]^{\mathbb{N}_{n}} \nonumber \\
      &\quad \cdot
      \left[
        1 - r^{(\cdot)}
      \right]^{\mathbb{N}_{P}\setminus\{i_{1},\ldots,i_{n}\}} \nonumber \\
      &\quad \cdot
      \left(
        \mathcal{R}(\emptyset; \overline{\mathcal{S}}(i_{1},... i_{n}))
        +
        \sum_{\ell=1}^{n}
        \hat{\mathcal{R}}_{\mathrm{z}}^{i_{\ell}}
      \right)
    \Bigg]
  \end{align}
  The above equation can be expressed in a more convenient form using disjoint index sets as
  \begin{equation}
    \mathrm{E}[\mathcal{R}] = \hspace{-0.4cm}
      \sum_{\mathcal{I}_{0} \uplus \mathcal{I}_{1} = \mathbb{N}_{P}} \!\!
      \left[r^{(\cdot)}\right]^{\mathcal{I}_{1}}
      \left[
        1 - r^{(\cdot)}
      \right]^{\mathcal{I}_{0}}
      \left[
        \sum_{i \in \mathcal{I}_{0}}
        \mathcal{R}(\emptyset; \overset{i}{\mathcal{S}})
        +
        \sum_{\ell \in \mathcal{I}_{1}}
        \hat{\mathcal{R}}_{\mathbf{z}}^{\ell}
      \right]
  \end{equation}
  Through the introduction of indicator functions, the summation hierarchy can be changed as follows:
  \begin{align}
    \mathrm{E}[\mathcal{R}]
    &=
    \sum_{j=1}^{P}
      \mathcal{R}(\emptyset; \overset{j}{\mathcal{S}})
      \sum_{\mathcal{I}_{0} \uplus \mathcal{I}_{1} = \mathbb{N}_{P}}
      \left[r^{(\cdot)}\right]^{\mathcal{I}_{1}}
      \left[
        1 - r^{(\cdot)}
      \right]^{\mathcal{I}_{0}}
       \cdot 1_{\mathcal{I}_{0}}(j)
      \nonumber\\
    &\qquad
      +
    \sum_{j=1}^{P}
      \hat{\mathcal{R}}_{\mathrm{z}}^{j}
      \sum_{\mathcal{I}_{0} \uplus \mathcal{I}_{1} = \mathbb{N}_{P}}
      \left[r^{(\cdot)}\right]^{\mathcal{I}_{1}}
      \left[
        1 - r^{(\cdot)}
      \right]^{\mathcal{I}_{0}}
      \cdot  1_{\mathcal{I}_{1}}(j)
      \label{eq:ExpectedRewardIndicatorFunctions}
  \end{align}
  Consider the first line of (\ref{eq:ExpectedRewardIndicatorFunctions}). The inner summand is only non-zero when $j\in\mathcal{I}_{0}$, so all non-zero terms share the common factor $(1 - r^{j})$.
  Similarly, in the second line, all non-zero terms in the inner summation share a common factor of $r^{j}$.
  Thus these terms are factored out, reducing the inner summation to one over disjoint subsets of $\mathbb{N}_{P}\setminus j$ as
  \begin{align}
    \mathrm{E}[\mathcal{R}]
    &=
    \sum_{j=1}^{P}
      \mathcal{R}(\emptyset; \overset{j}{\mathcal{S}})
      \left(1 - r^{j}\right)
      \sum_{\mathcal{I}_{0} \uplus \mathcal{I}_{1} = \mathbb{N}_{P} \setminus j}
      \left[r^{(\cdot)}\right]^{\mathcal{I}_{1}}
      \left[
        1 - r^{(\cdot)}
      \right]^{\mathcal{I}_{0}} \nonumber\\
    &\qquad
      +
    \sum_{j=1}^{P}
      \hat{\mathcal{R}}_{\mathrm{z}}^{j}
      \cdot
      r^{j}
      \sum_{\mathcal{I}_{0} \uplus \mathcal{I}_{1} = \mathbb{N}_{P} \setminus j}
      \left[r^{(\cdot)}\right]^{\mathcal{I}_{1}}
      \left[
        1 - r^{(\cdot)}
      \right]^{\mathcal{I}_{0}}
      \label{eq:ExpectedRewardExistenceFactored}
  \end{align}
  Manipulation of the inner sum yields
  \begin{align}
      &\sum_{\mathcal{I}_{0} \uplus \mathcal{I}_{1} = \mathbb{N}_{P} \setminus j}
      \left[r^{(\cdot)}\right]^{\mathcal{I}_{1}}
      \left[
        1 - r^{(\cdot)}
      \right]^{\mathcal{I}_{0}}\\
      &\qquad=
      \left[1 - r^{(\cdot)}\right]^{\mathbb{N}_{P}\setminus j}
      \sum_{\mathcal{I}_{0} \uplus \mathcal{I}_{1} = \mathbb{N}_{P} \setminus j}
      \left[\frac{r^{(\cdot)}}{1 - r^{(\cdot)}}\right]^{\mathcal{I}_{1}}
  \end{align}
  By the binomial theorem \cite[p.\ 369]{MahlerStatisticalMultitargetFusion07},
  \begin{align}
      \sum_{\mathcal{I}_{0} \uplus \mathcal{I}_{1} = \mathbb{N}_{P} \setminus j}
      \left[\frac{r^{(\cdot)}}{1 - r^{(\cdot)}}\right]^{\mathcal{I}_{1}}
      =
      \left[1 + \frac{r^{(\cdot)}}{1 - r^{(\cdot)}}\right]^{\mathbb{N}_{P} \setminus j}
  \end{align}
  Thus,
  \begin{align}
      &\sum_{\mathcal{I}_{0} \uplus \mathcal{I}_{1} = \mathbb{N}_{P} \setminus j}
      \left[r^{(\cdot)}\right]^{\mathcal{I}_{1}}
      \left[
        1 - r^{(\cdot)}
      \right]^{\mathcal{I}_{0}} \nonumber \\
      &\qquad=
      \left[
        \left(1 - r^{(\cdot)}\right)
        \left(1 + \frac{r^{(\cdot)}}{1 - r^{(\cdot)}}\right)
      \right]^{\mathbb{N}_{P} \setminus j}\\
      &\qquad=
      \left[1 - r^{(\cdot)} + r^{(\cdot)}\right]^{\mathbb{N}_{P}\setminus j}
      =1
  \end{align}
  With this, (\ref{eq:ExpectedRewardExistenceFactored}) simplifies to
  \begin{align}
    \mathrm{E}[\mathcal{R}]
    &=
    \sum_{j=1}^{P}
      \mathcal{R}(\emptyset; \overset{j}{\mathcal{S}})
      \left(1 - r^{j}\right)
      +
    \sum_{j=1}^{P}
      \hat{\mathcal{R}}_{\mathrm{z}}^{j}
      \cdot
      r^{j}
  \end{align}
  from which (\ref{eq:ExpectedRewardSum}) follows, completing the proof.
\section{Cell-Additivity of PHD-based KLD Information Gain}
\label{app:PhdKldIsAdditiveProof}
By (\ref{eq:NoCellOverlap}), the pseudo-likelihood can be written in terms of a sum of cell pseudo-likelihood functions
\begin{equation}
  L_{Z}(\mathbf{x}; \mathcal{S}) =
  \begin{cases}
    \sum_{j=1}^{P} 1_{\overset{j}{\mathcal{S}}}(\mathbf{s}) L_{Z}^{(j)}(\mathbf{x};\overset{j}{\mathcal{S}}) & \mathbf{s} \in \mathcal{S}\\
		1 & \mathbf{s}\notin\mathcal{S}
	\end{cases} \,.
  \label{eq:likelihood_as_sum}
\end{equation}
where
\begin{align}
  L_Z^{(j)}(\mathbf{x};\overset{j}{\mathcal{S}})  = \begin{cases} 1 - p_{D,j}(\mathbf{x}) + p_{D,j}(\mathbf{x})
    \Phi_{k}^{(j)}(Z|\mathbf{x}) & \mathbf{s} \in \overset{j}{\mathcal{S}} \\
    1  & \mathbf{s} \notin \overset{j}{\mathcal{S}}
  \end{cases} \, ,
\end{align}
\begin{equation}
	\Phi_k^{(j)} (Z|\mathbf{x}) = \sum_{\mathbf{z}\in Z \cap \overset{j}{\mathbb{Z}}}
	\frac{g_k(\mathbf{z}|\mathbf{x})}{\kappa_k(\mathbf{z}) +
	\int_{\overset{j}{\mathbb{X}}} D_{k|k-1}(\mathbf{x}')
  p_{D}(\mathbf{x}')
	g_k(\mathbf{z}|\mathbf{x}') d\mathbf{x}'};
  \label{eq:phi_def}
\end{equation}
and $p_{D,j}(\mathbf{x})=p_{D,k}(\mathbf{x}; \overset{j}{\mathcal{S}})$ is used for brevity.
Substituting (\ref{eq:likelihood_as_sum}) into (\ref{eq:KldPhdPseudolikelihood}) and noting that $L_{Z}^{(i)}(\mathbf{x}; \overset{j}{\mathcal{S}}) = L_{Z \cap \overset{i}{\mathbb{Z}}}^{(i)}(\mathbf{x};\overset{j}{\mathcal{S}})$,
\begin{align}
  \mathcal{R}_k(Z; \mathcal{S}) &=
  \sum_{j=1}^{P}\int D_{k|k-1}(\mathbf{x}) \bigg\{
    1 - L_{Z \cap \overset{j}{\mathbb{Z}}}^{(j)}(\mathbf{x}; \overset{j}{\mathcal{S}}) \nonumber \\
                                &+ L_{Z \cap \overset{j}{\mathbb{Z}}}^{(j)}(\mathbf{x}; \overset{j}{\mathcal{S}}) \log[L_{Z \cap \overset{j}{\mathbb{Z}}}^{(j)}(\mathbf{x}; \overset{j}{\mathcal{S}})]
	\bigg\} d\mathbf{x}
\end{align}
Thus, by comparing this form to that of (\ref{eq:KldPhdPseudolikelihood}), we see that the information gain is, in fact, a sum of information gains over the cells:
\begin{align}
  \mathcal{R}_k(Z; \mathcal{S}) &=
  \sum_{i=1}^{P} \mathcal{R}_{k}(Z \cap \overset{i}{\mathbb{Z}} ; \overset{i}{\mathcal{S}})
\end{align}

\section{Quadrature of Single-measurement Conditioned Information Gain Expectation}%
\label{app:Quadrature}
The quadrature approximation in (\ref{eq:HistogramQuadrature}) is most accurate when all measurement points within a given region yield a similar information gain.
However, performing excessive information gain computations to determine the quadrature regions would be counterproductive.
Instead, the regions $\{\overset{j}{\Omega}_{i}\}$ and their representative quadrature points $\mathbf{z}_{j,i}$ are selected using discrete samples of the predicted measurement \ac{phd}.

Let $\{\bar{\mathbf{z}}_{j}[\ell]\}_{\ell=1}^{Q}$ be an array of $Q\gg R_{j}$ uniformly spaced measurement samples on $\overset{j}{\mathbb{Z}}$ and
\begin{equation}
  \bar{D}_{j}[\ell] \triangleq D_{\mathrm{v},k|k-1}(\bar{\mathbf{z}}_{j}[\ell]; \mathcal{S})
\end{equation}
As illustrated in Fig.~\ref{fig:HistogramQuadrature}, the quadrature regions can be represented by sets of measurement points with similar log-\ac{phd} values; i.e.,
\begin{align}
  \overset{j}{\Omega}_{i}
  =
  \begin{cases}
    \{\bar{\mathbf{z}}_{j}[\ell] \, : \, 1\leq \ell \leq Q, \, \bar{D}_{j}[\ell] \leq e^{\varepsilon_{j,i}}\} & i=1 \\
    \{\bar{\mathbf{z}}_{j}[\ell] \, : \, 1\leq \ell \leq Q, \, e^{\varepsilon_{j,i-1}}< \bar{D}_{j}[\ell] \leq e^{\varepsilon_{j,i}}\} & i>1 \\
  \end{cases}
\end{align}
where the discrete logarithmic bin edges are obtained as
\begin{align}
  \varepsilon_{j,i}
  &=
  \varepsilon_{0} + \frac{i}{R_{j}}(\varepsilon_{R_{j}} - \varepsilon_{0}), \quad 0<i<R_{j} \\
  \label{eq:QuadratureLogLowerBound}
  \varepsilon_{j,0}
  &=
  \max
  \left[
    \varepsilon_{\min},
    \min_{1<\ell< Q}
    \left(
      \log\{\bar{D}_{j}[\ell]\}
    \right)
  \right] \\
  \varepsilon_{j,R_{j}}
  &= \max_{1\leq \ell \leq Q}
  \left(
    \log\{\bar{D}_{j}[\ell]\}
  \right)
\end{align}
In (\ref{eq:QuadratureLogLowerBound}), the tunable parameter $\varepsilon_{\min}$ represents the lowest log-\ac{phd} that should be considered, so as to reduce unnecessary information gain computations in areas of extremely low probability.
Then, a representative measurement is chosen from each region as
\begin{equation}
  \label{eq:QuadratureRepresentativeMeasurement}
  \mathbf{z}_{j,i}
  =
  \argmin_{\bar{\mathbf{z}}[\ell] \in \overset{j}{\Omega}_{i}}
  \left[
    | \bar{D}_{j}[\ell]
    - \hat{D}_{j,i} |
  \right]
\end{equation}
where, in (\ref{eq:QuadratureRepresentativeMeasurement}), $|\cdot|$ represents the absolute value operator, and $\hat{D}_{j,i}$ is the average \ac{phd} value in region $i$ of measurement cell $j$:
\begin{align}
  \hat{D}_{j,i} = \frac{1}{|\overset{j}{\Omega}_{i}|}
  \sum_{1\leq\ell \leq Q, \, \bar{\mathbf{z}}_{j}[\ell] \in \overset{j}{\Omega}_{i}} \bar{D}_{j}[\ell]
\end{align}
The volumes can be approximated by the proportion of discrete measurement samples that fall within each region as
\begin{align}
  A_{j,i} = \frac{| \overset{j}{\Omega}_{i}|}{Q} \cdot A(\overset{j}{\mathbb{Z}})
\end{align}

\ifCLASSOPTIONcaptionsoff
  \newpage
\fi

\end{document}